\documentclass[aps,onecolumn,11pt,floatfix,altaffilletter,preprintnumbers,tightenlines,showpacs,showkeys,notitlepage,nofootinbib]{revtex4-1}
\pdfoutput=1
\usepackage[colorlinks=true,citecolor=blue,linkcolor=blue]{hyperref}
\usepackage[normalem]{ulem}
\usepackage{mathtools}
\usepackage{amsmath,amssymb}
\usepackage{array}
\usepackage{epsfig}
\usepackage{slashed}
\usepackage{graphicx}               
\usepackage{url}
\usepackage{color}
\usepackage{multirow}
\usepackage{booktabs}
\usepackage{placeins}
\usepackage[dvipsnames]{xcolor}
\usepackage{epstopdf}
\usepackage{verbatim}
\usepackage{tikz}
\usetikzlibrary{trees}
\usetikzlibrary{decorations.pathmorphing}
\usetikzlibrary{decorations.markings}
\usepackage{bbold}
\usepackage{cancel}
\definecolor{jblue}  {RGB}{20,50,100}
\definecolor{npurple}  {RGB} {153, 51, 204}
\definecolor{wred}   {RGB}{217,0,56}
\definecolor{white}   {RGB}{255,255,255}

\definecolor{korange}   {RGB}{235, 80,  43}
\definecolor{korange2}   {RGB}{245, 100,  63}
\definecolor{kyelloworange}   {RGB}{255, 210,  110}
\definecolor{kyelloworange2}   {RGB}{240, 170,  90}
\definecolor{kred}   {RGB}{204,  102, 153}
\definecolor{kpurple}   {RGB}{153,  61, 190}
\definecolor{kpurplelight}   {RGB}{213,  161, 230}

\usepackage{siunitx}

\usepackage[capitalize]{cleveref}
\usepackage{subfigure} 
\usepackage{lipsum}
\definecolor{red}{rgb}{1.0, 0, 0}
\usepackage{gensymb}
\usepackage{feynmp-auto}
\allowdisplaybreaks

\bibpunct{[}{]}{,}{n}{}{,}

\newcolumntype{C}{>{$}c<{$}}
\AtBeginDocument{
\heavyrulewidth=.08em
\lightrulewidth=.05em
\cmidrulewidth=.03em
\belowrulesep=.65ex
\belowbottomsep=0pt
\aboverulesep=.4ex
\abovetopsep=0pt
\cmidrulesep=\doublerulesep
\cmidrulekern=.5em
\defaultaddspace=.5em
}

\newcommand{\nocontentsline}[3]{}
\newcommand{\tocless}[2]{\bgroup\let\addcontentsline=\nocontentsline#1{#2}\egroup}

\newcommand{\BR}{\text{BR}}

\newcommand{\keV}{\,\mathrm{keV}}
\newcommand{\MeV}{\,\mathrm{MeV}}
\newcommand{\GeV}{\,\mathrm{GeV}}
\newcommand{\TeV}{\,\mathrm{TeV}}
\newcommand{\dd}{\mathrm{d}}
\newcommand{\invfb}{\,\mathrm{fb}^{-1}}
\newcommand{\invab}{\,\mathrm{ab}^{-1}}
\newcommand{\fb}{\,\mathrm{fb}}

\newcommand{\met}{$\mathbf{+\cancel{E_T}}\;$}
\newcommand{\MET}{$\cancel{E_T}$\ }


\begin{document}

\title{Shining Light on the Scotogenic Model: \\ \vspace{0.1 cm} Interplay of Colliders and Cosmology} 
         
\author{Sven Baumholzer$^{1}$}            
\author{Vedran Brdar$^{2}$}	       
\author{Pedro Schwaller$^{1}$\,}           
\author{Alexander Segner$^{1}$\,}           
\affiliation{$^1$PRISMA$^+$ Cluster of Excellence and
             Mainz Institute for Theoretical Physics,
            Johannes Gutenberg-Universit\"{a}t Mainz, 55099 Mainz, Germany\\
$^2$Max-Planck-Institut f\"ur Kernphysik, Saupfercheckweg 1,
          69117 Heidelberg, Germany            }

\preprint{MITP/19-088}

\begin{abstract}
\noindent
In the framework of the scotogenic model, which features radiative generation of neutrino masses, we explore light dark matter scenario. Throughout the paper we chiefly focus on keV-scale dark matter which can be produced either via freeze-in through the decays of the new scalars, or from the decays of next-to-lightest fermionic particle in the spectrum, which is produced through freeze-out. The latter mechanism is required to be suppressed as it typically produces a hot dark matter component. Constraints from BBN are also considered and in combination with the former production mechanism they impose the dark matter to be light. For this scenario we consider signatures at High Luminosity LHC and proposed future hadron and lepton colliders, namely FCC-hh and CLIC, focusing on searches with two leptons and missing energy as a final state. While a potential discovery at High Luminosity LHC is in tension with limits from cosmology, the situation greatly improves for future colliders. 
\end{abstract}

\maketitle
{\hypersetup{linkcolor=black}
\tableofcontents}
\section{Introduction}
\label{sec:intro}
\noindent
While more than two decades have passed since the groundbreaking discovery of neutrino oscillations, which unambiguously established that the most elusive Standard Model (SM) particles are massive, the origin of neutrino mass still remains unknown. In spite of the viable scenario in which, by supplementing SM left-handed neutrino fields with right-handed components, neutrino masses are generated in the same way as for all the other fermions, the smallness of  Yukawa couplings required for generating eV-scale masses has led to a much greater interest in Majorana mass models. The famous realization of the latter possiblity is the type-I seesaw model \cite{Goran,Minkowski,GellMann:1980vs,Yanagida:1979as} in which neutrino masses are generated at tree-level  in the presence of at least two generations of heavy neutral leptons.
For ``natural" $\mathcal{O}(1)$ values of Yukawa couplings, this model suggests that the mass scale of heavy leptons is around $10^{13}$ GeV, clearly unreachable at any terrestrial experiment. In contrast, radiative neutrino mass models can lower the scale of new physics by several orders of magnitude.

Among radiative neutrino models, one of the simplest realizations is the so called ``scotogenic" model \cite{Ma:2006km}. Since a $\mathbf{Z}_2$ symmetry needs to be imposed in order to forbid the tree-level neutrino mass generation, the lightest among the newly introduced particles can be a viable dark matter (DM) candidate \cite{Molinaro:2014lfa,Lindner:2016kqk,Kumericki:2012bh,Brdar:2013iea,Avila:2019hhv,Picek:2019yga}. The success of thermal leptogenesis in this model has also been demonstrated \cite{Kashiwase:2012xd,Suematsu:2011va,Hugle:2018qbw,Racker:2013lua,Mahanta:2019gfe,Borah:2018rca,Mahanta:2019sfo} and, in addition, the authors of this work have recently shown that light DM and ARS leptogenesis \cite{Akhmedov:1998qx} can be embedded simultaneously in the framework of the $\nu\nu$MSM model \cite{Baumholzer:2018sfb}. It was found that one can have the spectrum of new particles below the TeV-scale, while bounds from cosmology require the mass of the light DM to be below $\mathcal{O}(10\keV)$. This raises the question whether one can test this model setup at present and future colliders, where the accessible energy range exceeds the masses of all newly introduced particles. Moreover, the light DM candidate can lead to interesting consequences for the early Universe and in this paper we put a special emphasis on a synergy between collider searches and cosmological probes for testing the scotogenic model.    

The paper is organized as follows. In \cref{sec:model} we introduce the 
particle content of the model and discuss the mechanism for the generation of neutrino masses. In \cref{sec:DM} we discuss the production of light fermionic DM and derive respective limits from cosmology. In \cref{sec:pp-collider}, for such a scenario with light DM and hierarchical spectrum in $\mathbf{Z}_2$-odd sector, we present the calculated projections for High Luminosity LHC (HL-LHC) search with two tau leptons or electrons/muons and missing transverse energy in the final state.
Cosmological bounds put the parameter space testable at HL-LHC in tension, which motivates us to go beyond and consider future colliders. In \cref{sec:future} we therefore present the discovery potential at FCC-hh \cite{Mangano:2017tke} and CLIC \cite{deBlas:2018mhx} by considering di-lepton +$\cancel{E}_T$ channel. 
 We conclude in \cref{sec:conclusion}.

\section{The Model and Neutrino Masses}
\label{sec:model}
\noindent
We consider the scotogenic model which, in addition to the SM field content, contains one scalar doublet $\Sigma=(\sigma^+, \sigma^0)^\text{T}$ as well as three generations of heavy neutral leptons (HNLs) $N_i\, (i=1,2,3)$. In addition to these novel fields, the model requires a discrete $\mathbf{Z}_2$ symmetry under which new degrees of freedom have an odd charge. The part of the Lagrangian containing newly introduced fields is 
\begin{align}
\mathcal{L}\supset \frac{i}{2} \bar{N}_i\, \slashed{\partial}\, N_i -\left(y_{i\alpha}\, \bar{N}_i\, \tilde{\Sigma}^\dag L_\alpha+\frac{1}{2} m_{N_i}\bar{N_i} N_i^c+\text{h.c.}\right)+ (D_\mu \Sigma)^\dag (D^\mu \Sigma) -V(\Phi,\Sigma)\,,
\label{eq:lagrangian}
\end{align}
where $y_{i\alpha}$ is the Yukawa coupling between $i$-th HNL, $\Sigma$ and SM lepton doublet $L_\alpha=(\nu_\alpha, \alpha^-)^\text{T}\,(\alpha=e,\,u,\tau)$, $m_{N_{i}}$
is the mass of i-th HNL, $D_\mu$ is the covariant derivative, $\Phi=(\phi^+,\phi^0)^\text{T}$ is the SM Higgs doublet, and $V(\Phi,\Sigma)$ represents the scalar potential
\begin{align}
V(\Phi,\Sigma)=&\mu_{1}^2\, \Phi^\dag \Phi +\mu_{2}^2\, \Sigma^\dag \Sigma+\frac{1}{2}\,\lambda_1\, (\Phi^\dag \Phi)^2+\frac{1}{2}\,\lambda_2 \,(\Sigma^\dag \Sigma)^2 +\lambda_3\,(\Phi^\dag \Phi)(\Sigma^\dag \Sigma)\nonumber \\&+\lambda_4\,(\Phi^\dag \Sigma)(\Sigma^\dag \Phi)+\frac{\lambda_5}{2}\,\left((\Phi^\dag \Sigma)^2+\text{h.c.}\right).
\label{eq:potential}
\end{align}
The couplings in the scalar sector are constrained from the vacuum stability requirement \cite{Branco:2011iw,Lindner:2016kqk}
\begin{align}
\lambda_3 &>-\sqrt{\lambda_1 \lambda_2}\,, & \lambda_3+\lambda_4-|\lambda_5|&> -\sqrt{\lambda_1 \lambda_2}\,, & \lambda_{1,2}&>0\,.
\label{eq:couplings_constants_boundaries}
\end{align}
From \cref{eq:potential}, we can directly infer the masses of novel scalar degrees of freedom after electroweak symmetry breaking (EWSB)
\begin{align}
 &m_{\pm}^2 =  \mu_2^2 + \lambda_3 v^2, \nonumber \\ &
  m_S^2 = \mu_2^2+(\lambda_3+\lambda_4+\lambda_5)\,v^2, \nonumber \\ &
   m_A^2 = \mu_2^2+(\lambda_3+\lambda_4-\lambda_5)\,v^2\,,
   \label{eq:ms}
\end{align}
where $v=246/\sqrt{2}$ is the vacuum expectation value of the Higgs field. In the first line of \cref{eq:ms} the mass of charged scalars is given, whereas the latter two masses correspond to the CP-even (S) and CP-odd (A) neutral scalars, defined as $\sigma^0 = (S + iA)/\sqrt{2}$.

Since the exact $\mathbf{Z}_2$ symmetry forbids tree-level neutrino masses, they are realized radiatively with the following expression obtained by calculating self-energy corrections to the neutrino propagator from the exchange of neutral spin-zero  $S$ and $A$ fields \cite{Ma:2006km,Toma:2013zsa,Merle:2015ica}
\begin{align}
 (m_\nu)_{\alpha\beta} = \sum\limits_i \frac{y_{i\alpha}y_{i\beta}m_{N_i}}{32\pi^2} \left[ \frac{m_S^2}{m_S^2-m_{N_i}^2}\mathrm{ln}\left(\frac{m_S^2}{m_{N_i}^2}\right) -
 \frac{m_A^2}{m_A^2-m_{N_i}^2}\mathrm{ln}\left(\frac{m_A^2}{m_{N_i}^2}\right) \right]\equiv \sum_i y_{i\alpha} y_{i\beta}\, \Lambda_i \,.
 \label{eq:radiative_masses}
\end{align}
Here, the summation index runs over HNL generations and in the last equality we abbreviated this formula, modulo Yukawa couplings, with $\Lambda_i$. In the present work, the mass of the lightest HNL is $\mathcal{O}(\text{keV})$ with $y_{1\alpha}\simeq \mathcal{O}(10^{-8})$ and this state effectively does not participate in the neutrino mass generation (\ref{eq:radiative_masses}). This makes the lightest active neutrino effectively massless, which is a viable scenario, consistent with the data from neutrino oscillation experiments that are probing only  mass squared differences. In this case, $N_1$ is decoupled from the mass generation and, hence, only the elements of $2\times3$ submatrix of $y$, $y_{23}$, enter in \cref{eq:radiative_masses}.

In order to properly account for low-energy neutrino data in the analysis, we employ the Casas-Ibarra parametrization \cite{Casas:2001sr} which imposes the following expression for the Yukawa submatrix 
\begin{align}
 y_{23}= \mathrm{i} \left(\sqrt{\Lambda^\text{diag}}\right)^{-1}R\,\sqrt{m_\nu^\text{diag}}\,U_\mathrm{PMNS}^\dagger\,,
 \label{eq:Casas-Ibara-Para}
\end{align}
where $\Lambda^\text{diag}=\text{diag}(\Lambda_2,\Lambda_3)$, and $R$ is an orthogonal matrix parametrized with a complex angle $\vartheta=\omega-\mathrm{i}\,\eta$
\begin{equation}
  R(z)=\begin{cases}
    \begin{pmatrix}
    0 & \cos \vartheta & -\sin \vartheta \\
    0 & \sin \vartheta & \cos \vartheta
    \end{pmatrix}\,, & \text{for normal neutrino mass ordering (NO)}\,,\\
    \begin{pmatrix}
     \cos \vartheta & -\sin \vartheta & 0\\
     \sin \vartheta & \cos \vartheta & 0
    \end{pmatrix}\,, & \text{for inverted  neutrino mass ordering (IO)}\,.
  \end{cases}
  \label{eq:CIR}
\end{equation}

The remaining ingredients in \cref{eq:Casas-Ibara-Para} are the neutrino masses

\begin{equation}
   m_\nu^\text{diag}=
   \begin{cases}
    \text{diag}\left(0\,,\sqrt{m_\mathrm{sol}^2},\sqrt{m_\mathrm{atm}^2}\right)\,, & \text{for NO}\,,\\
    \text{diag}\left(\sqrt{m_\mathrm{atm}^2},\sqrt{m_\mathrm{sol}^2+m_\mathrm{atm}^2},\,0 \right)\,, & \text{for IO}\,,
  \end{cases}
\end{equation}
where $m_\mathrm{sol}^2$ and $m_\mathrm{atm}^2$ are solar and atmospheric mass squared differences, and the leptonic mixing matrix, $U_\text{PMNS}$, which is parameterized as in Ref.~\cite{Patrignani:2016xqp}.
The relevant parameters for us are one Dirac $(\delta)$ and two Majorana CP phases $(\alpha_1, \alpha_2)$. While the mixing angles are relatively precisely determined (see \cite{Esteban:2018azc} for recent NuFIT results that we employ in this work), the value of the Dirac CP phase is practically unconstrained, and Majorana phases are not testable at neutrino oscillation facilities.

The elements of Yukawa couplings in $y_{23}$ are constrained from above 
due to non-observation of LFV processes such as $\alpha\to \alpha'\gamma$ and $\alpha\to 3 \alpha'$ where $\alpha$ and $\alpha'$ denote different species of charged leptons. The upper bounds on the branching ratios (BR) for these types of decays are given in \cite{Patrignani:2016xqp} and also compiled in Table 1 in \cite{Baumholzer:2018sfb}. While we have implemented all available constraints from LFV decays it is worthwhile pointing out that the dominant effect arises from the lack of observation of $\mu\to e \gamma$ process. The upper bound on the BR for this process is $4.2\times 10^{-13}$ which converts to \cite{Toma:2013zsa}
\begin{align}
\Big|\sum_{i=2,3} y_{i\mu} \, y_{ie}^*\Big| \lesssim 4.3\times 10^{-3}\,\left(\frac{m_\pm}{1\,\text{TeV}}\right)^2,
\label{eq:LFV}
\end{align}
for $m_{N_{2,3}}\simeq 0.1$ TeV. 

Finally, the neutrino mass matrix in \cref{eq:radiative_masses} depends on $\lambda_5$ which enters in the expression for $m_S$ and $m_A$ (see again \cref{eq:ms}).
This formula actually features a linear dependence of neutrino masses on $\lambda_5$ for small values of $\lambda_5$ \cite{Ma:2006km}.
Parameters $\lambda_5$ and entries of $y_{23}$ depend on each other, and jointly set the scale for neutrino mass $\sim 0.1$ eV. This means that there is a lower bound  $\lambda_5 \gtrsim \mathcal{O}(10^{-7})$, calculated for $m_\pm=\mathcal{O}(1)$ TeV. Constraints arising from electroweak precision data \cite{Ahriche:2017iar} are not competitive to the above discussed ones.

\section{Dark Matter and Cosmology}
\label{sec:DM}
\noindent
Since the lightest of the newly introduced particles is stable, it is natural to consider whether it can account for the DM in the Universe. In the scotogenic model there are neutral particles both in the fermionic and scalar sector, making them potential candidates. Motivated by our previous work \cite{Baumholzer:2018sfb} we stick to fermionic DM and keep the scalars heavier than HNLs.
It was shown in Refs. \cite{Molinaro:2014lfa,Baumholzer:2018sfb} that in the scotogenic setup with fermionic $\mathcal{O}(100)$ GeV DM, the relic abundance from freeze-out generally strongly overshoots the measured values. There are, however, options how to remedy this problem:\\

\noindent
$(i)$ If additional processes, namely coannihilations of DM with new scalars, are involved, DM can stay longer in the thermal equilibrium and freeze-out with much smaller abundance (see for instance Fig.~2 in \cite{Baumholzer:2018sfb}). The coannihilations are only effective if the  splitting between the DM and scalar mass is tiny. Such a scenario would yield soft final-state leptons which are hard to reconstruct when considering di-lepton and di-tau searches (see \cref{sec:pp-collider,sec:future}) and hence freeze-out of $\mathcal{O}(100)$ GeV DM is not compatible with the signatures at hadron colliders that are studied in this paper. This conclusion changes for the case of the future lepton collider CLIC, for which we will show in \cref{sec:future} that such regime can be probed as well\footnote{Let us note that the realization with a very compressed spectrum of $\mathbf{Z}_2$-odd particles \cite{Schmidt:2012yg,Modak:2014vva,Faisel:2014gda,Biswas:2017ait} is also testable using astrophysical observations \cite{Abazajian:2006jc,Bulbul:2014sua,Boyarsky:2014jta,Cappelluti:2017ywp,Brdar:2017wgy,Caputo:2019djj}. This would in particular require small splitting in $\mathbf{Z}_2$-odd fermion sector, scenario which will not be further explored in this work.}. \\

\noindent
$(ii)$ The overproduction problem can be solved by considering light, non-thermally produced DM. As we have shown in \cite{Baumholzer:2018sfb}, such DM can be produced either via freeze-in \cite{Hall:2009bx, Kusenko:2006rh} from the decays of neutral and charged scalars in the $\Sigma$ doublet or from the decays of frozen-out next-to-lightest HNL, \emph{i.e.} $N_2$ \footnote{Such a scenario was already considered in \cite{Molinaro:2014lfa}.}. However, the latter mechanism is also constrained by requiring $N_2$ to decay before the time of big bang nucleosynthesis (BBN). Namely, if $N_2$ is too long lived, the abundances of light nuclei will be altered. This production mechanism also leads to too hot momentum distributions and therefore needs to be subdominant with respect to scalar decay contribution. We elaborate on this in the present section. 

Generally, in the freeze-in scenario there is some freedom to choose a DM mass. On the contrary, in this model there is an upper bound from demanding that the DM production has stopped before BBN takes place, as will be discussed in the last part of this section. We would like to point out, that one can live with DM masses up to few\,MeV and in that case limits from structure formation, discussed in the second part of this section, are weakened. However, in the following we will consider the DM mass to be $\mathcal{O}(10\keV)$. This particular choice is motivated by our findings in \cite{Baumholzer:2018sfb} where we identified an interesting open window in parameter space to successfully incorporate leptogenesis.


\subsection{Production mechanisms}
\label{subsec:production}
\noindent
The processes through which keV-scale $N_1$ is frozen-in are
\begin{align}
A,S\to N_1 \,\nu_\alpha\,,\hspace{5mm}\sigma^\pm\to N_1\, l^\pm_\alpha\,.
\label{eq:decay}
\end{align}
The corresponding Boltzmann equation for the DM yield, $Y_\text{FI}$,  which is the ratio of DM number density and entropy density, reads \cite{Baumholzer:2018sfb,Hall:2009bx} 
\begin{align}
\frac{dY_{\text{FI}}}{dx}=\frac{135\, M_{\text{Pl}}\, |y_1|^2}{1.66\cdot 64\,\pi^5 g_*^{3/2} m_\pm} x^3 \bigg(2\,K_1(x)+r_A^3  \,K_1(r_A \, x) + r_S^3\, K_1(r_S\, x) \bigg)\,.
\label{eq:yield-diff-dx}
\end{align}
Here, $M_{\text{Pl}} =1.22\times 10^{19}\GeV$ is the Planck mass, and the number of relativistic degrees of freedom across relevant temperatures is fixed to $g_*=114.25$, taking new particles into account. In \cref{eq:yield-diff-dx} it is assumed for simplicity that the Yukawa couplings of $N_1$ are flavor universal, \emph{i.e.} $y_{1\alpha}\equiv y_1$. Furthermore, the following abbreviations  $r_A=m_A/m_\pm$ and $r_S=m_S/m_\pm$ are introduced to account for all three production channels. $K_1$ is the modified Bessel function of the second kind and the redefined temperature  $x=m_\pm/T$ is employed. In the computation, we adopted Maxwell-Boltzmann distributions for phase space densities of all thermalized species involving $\Sigma$. Finally, we assumed the initial DM number density to vanish and this simplifies the computation of DM abundance. Namely, instead of solving differential equation, a straightforward integration suffices.\\

We can obtain the expression for $Y_\text{FI}$ by simply integrating 
\cref{eq:yield-diff-dx} between the temperature at the end of inflation and the present one, where the former is associated to the reheating temperature which is assumed to be larger than all particle masses in the model. Practically, this allows us to use $x=0$ and $x=\infty$ as the respective integration boundaries. We obtain
\begin{align}
Y_\text{FI}= \frac{405 M_\text{Pl}\, |y_1|^2}{128\pi^4\cdot1.66\cdot g_*^{3/2} m_\pm}\frac{2 r_A r_S +r_S+r_A}{r_A r_S}\,.
\label{eq:yield-integrated}
\end{align}
By taking $\lambda_4=\lambda_5=0$ which renders $r_S=r_A=1$ and by using the relation between yield and relic abundance, $\Omega h^2 =2.742\cdot 10^2 \,\left(m_{N_1}/\text{keV}\right) \,Y_\text{FI}$, we arrive at the analytical estimate for DM relic abundance 
\begin{align}
\Omega h^2_\text{FI} \approx 0.12\, \bigg(\frac{|y_1|}{2.36\cdot10^{-8}}\bigg)^2 \bigg(\frac{m_{N_1}}{1\, \text{keV}} \bigg)\bigg(\frac{1\TeV}{m_{\pm}} \bigg)\,,
\label{eq:relic_abundance}
\end{align}
from which one infers that in order to have scalar decays as a dominant DM production mechanism, the required DM Yukawa couplings need to be $\mathcal{O}(10^{-8})$ for $\mathcal{O}(\text{TeV})$ masses of new scalars and keV-scale $N_1$.

In addition to the described freeze-in mechanism, DM in this model can be produced from the decays of next-to-lightest $\mathbf{Z}_2$-odd particle, $N_2$. The Yukawa couplings $y_{2\alpha}$, that are required for the successful generation of neutrino masses via mechanism described in \cref{sec:model}, are sufficiently strong to put this particle in thermal equilibrium with the SM bath. Hence, $N_2$ will freeze-out at $y=m_{N_2}/T\lesssim 15$\,. Note that the $\mathbf{Z}_2$-odd scalars are 
also in thermal equilibrium due to gauge interactions. Still, all such particles eventually decay to $N_2$ and hence one effectively needs to solve  a single Boltzmann equation for $N_2$  
\begin{align}
\frac{dY}{dy}=\sqrt{\frac{\pi g_*}{45}}\, \frac{M_\text{Pl} \, m_{N_2}}{y^2}\, \langle \sigma_\text{eff} v \rangle \,(Y_\text{eq}^2-Y^2)\,,
\label{eq:freeze-out}
\end{align}    
where $\langle \sigma_\text{eff} v \rangle$ accounts for the annihilations and coannihilations in the $\mathbf{Z}_2$-odd sector. The multitude of relevant processes enforces the evaluation of \cref{eq:freeze-out} with numerical tools and to this end we employed \texttt{micrOMEGAs\,5.1} \cite{Barducci:2016pcb}. For a  detailed description of our implementation as well as the procedure to derive \cref{eq:freeze-out} we refer the reader to our previous publication \cite{Baumholzer:2018sfb} where we also demonstrated the strong effect of coannihilations to the freeze-out abundance of $N_2$. 

After freeze-out, $N_2$ decays into $N_1$ and a pair of charged or neutral leptons with the rate \cite{Molinaro:2014lfa}

\begin{align}
\Gamma (N_2 \to l_\alpha l_\beta N_1) =\frac{m_{N_2}^5}{6144 \,\pi^3 M^4} \bigg(|y_{1\beta}|^2 |y_{2\alpha}|^2 + |y_{1\alpha}|^2 |y_{2\beta}|^2 \bigg)\,,
\label{eq:decay_N2}
\end{align}
where $M$ stands for the mass of the scalar particle that is exchanged in the process and $\alpha$ and $\beta$ denote the flavor of final state leptons.

The decay of $N_2$ gives a contribution to the total DM abundance of the form

\begin{align}
\Omega h^2_{N_2\to N_1} = \frac{m_{N_1}}{m_{N_2}}\, \Omega h^2_{N_2}\,,
\label{eq:freeze_out_relic}
\end{align}
where $\Omega h^2_{N_2}$ is the freeze-out abundance of $N_2$ that can be related to the corresponding yield, $Y$, calculated by solving \cref{eq:freeze-out}. Even though $N_2$ decays give an extra source of DM, this production mechanism actually yields two ``problems":

\begin{itemize}
\item $N_2$ decays occur after freeze-out, at temperatures much lower than $m_{N_1}$ and this leads to the production of  DM particles with a hot momentum distribution \cite{Heeck:2017xbu}. This is not a generic property of the model but it does occur for the parameter choices under our consideration. This can drastically suppress the structure at small scales and may not be compatible with observations. 
 
\item  The decays of $N_2$ should be fast enough in order not to violate  BBN predictions. For $N_2$ decays to $\tau$ leptons, which dominantly decay hadronically, the decay time needs to be $\tau_{N_2\to N_1}\lesssim 1$ sec, whereas decays into leptons of first and second generation do not lead to such stringent limits, yielding $\tau_{N_2\to N_1}\lesssim 100$ sec \cite{Kawasaki:2017bqm}. 
\end{itemize}


\subsection{Constraints from structure formation}
\label{subsec:Lyalpha}
\noindent
In this section we consider the compatibility of structure formation with the two previously described DM production mechanisms.  
The structure formation limits on keV-scale sterile neutrino DM are commonly derived for non-resonant production, dubbed Dodelson-Widrow \cite{Dodelson:1993je}, for which non-zero mixing between active and sterile states is required. Currently, the most stringent structure formation limit on the mass of non-resonantly produced particles $(m_\text{NRP})$ arises from Lyman-$\alpha$ forest data \cite{Yeche:2017upn} and yields $m_\text{NRP}\gtrsim 28.8$\,keV. This limit should be taken with a grain of salt because the Lyman-$\alpha$ forest absorption spectra can be dominated by the effects stemming from the gas dynamics in the Inter-Galactic Medium \cite{Kulkarni:2015fga}.  
More robust constraints arise from Milky Way satellite counts \cite{Merle:2015vzu} and give $m_\text{NRP}\gtrsim 10$\,keV. 
In order to derive constraints from these observations for our model, we evaluate the DM momentum distribution function $f_{N_1}(z,r)$ which is calculated as a function of the dimensionless variables $z\equiv p/T$ and $r\equiv m_P/T$. Here, $m_P$ stands for the mass of a parent particle, which are either heavy scalars in the case of freeze-in or $N_2$ in late time next-to-lightest particle decays. For scalar decays we are following the discussion in Ref. \cite{Heeck:2017xbu}, whereas for the case of $N_2$ decays we employ the procedure outlined in Ref. \cite{Merle:2015oja}.
The total distribution function is hence given as a sum of the two contributions
\begin{align}
    f_{N_1}(z,r) = f_{N_1}^\Sigma(z,r) +  f_{N_1}^{N_2}(z,r)\,,
\end{align}
where with superscripts we indicated the production mechanisms of $N_1$. In what follows we discuss the calculation of both components.\\
The general expression for $f_{N_1}^\Sigma(z,r)$ 
 assuming a Maxwell-Boltzmann distribution is given by \cite{Heeck:2017xbu,Merle:2015oja}:
\begin{align}
    f_{N_1}^\Sigma(z,r)= 4\,C_\Gamma^\Sigma \left( \frac{e^{-z}\sqrt{\pi}\,\text{Erf}\left[ \frac{r}{\sqrt{4z}} \right]}{2\sqrt{z}} - e^{-z\left( \frac{r^2}{4z^2}+1 \right)}\frac{r}{2z}\right) \;\underset{r\to \infty}{\longrightarrow}\; 4\,C_\Gamma^\Sigma \sqrt{\frac{\pi}{z}} e^{-z},
    \label{eq:fN1_scalar}
\end{align}
where $C_\Gamma^\Sigma=M_0\, \Gamma_\Sigma/m_\pm^2$  is the effective decay width, originating from rescaled 2-body decays of the $\Sigma$ particles
\begin{align}
    C_\Gamma^\Sigma = \frac{M_0}{m_\pm^2} \left( \frac{6|y_1|^2m_\pm}{16\pi} + \frac{3|y_1|^2m_S}{16\pi} +\frac{3|y_1|^2m_A}{16\pi} \right)\,,
    \label{eq:c_2body}
\end{align}
with $M_0\approx 7.35\, g_*^{-1/2} \times 10^{18}\GeV$.

From $f_{N_1}^\Sigma(z,r)$ we find the average DM momentum $\langle z \rangle_{\text{FI}}^\text{prod}\approx 2.5$, in agreement with eq.~(19) in \cite{Heeck:2017xbu}. This result, together with the information that the production dominantly occurs at temperatures $T\sim m_\Sigma /3$ (see for instance fig. 1 in \cite{Baumholzer:2018sfb}), allows us to estimate the limit on $m_{{N_1}}$ by using \cite{Heeck:2017xbu}
\begin{align}
m_{{N_1}} =\frac{{\langle p/T \rangle}^\text{prod}}{3.15}\,\left(\frac{10.75}{g_*(T^\text{prod})}\right)^{1/3}\, m_\text{NRP}\,,
\label{eq:limit-cosmo}
\end{align}
and assuming that the freeze-in DM production dominates. Here, the entropy dilution factor $(10.75/g_*(T^\text{prod}))^{1/3}$ takes into account that the DM production happens at early times.
Taking the aforementioned limit $m_\text{NRP}\gtrsim 10\keV$ we obtain $m_{N_1}>3.7\keV$. 
The combination of this limit and \cref{eq:relic_abundance} sets the upper bound on the values of $y_{1\alpha}$. If decays of $\mathbf{Z}_2$-odd scalars were the only source of DM production, our structure formation analysis would end here. However, decays of $N_2$ significantly complicate the picture. 
To calculate the DM distribution function for the production via $N_2$ decays,
we apply the \emph{master equation}\footnote{This equation was derived for 2-body decays, while $N_2$ decays into three particles, including $N_1$.  The derivation of such a general expression for 3-body decays is beyond the scope of the present work. However, note that by considering formula for 2-body decays we are actually being conservative because in such case DM is typically emitted with larger momentum with respect to the realistic 3-body case.
} from \cite{Merle:2015oja} and evaluate it numerically

\begin{align}
f^{N_2}_{N_1}(z,r)=\int_{r_\text{FO}}^r \dd r'\, \,C_\Gamma^{N_2}\, \frac{r'^2}{z^2} \int_{|z-r'^2/(4z)|}^\infty \dd\hat{z} \frac{\hat{z}}{\sqrt{\hat{z}^2+r'^2}}\, f_{N_2}(\hat{z},r')\,.
\label{eq:master}
\end{align}
Here, $r_\text{FO} \in (8,\,16)$ is evaluated at the freeze-out temperature and $C_\Gamma^{N_2}$ is the effective decay width given by $C_\Gamma^{N_2}=M_0\, \Gamma/m_{{N_2}}^2$, where $\Gamma$ is the decay width of $N_2$ into $N_1$ and a pair of leptons given in \cref{eq:decay_N2}. The expression for the distribution function of $N_2$ after freeze-out is \cite{Merle:2015oja}
\begin{align}
f_{N_2}(z,r)&=\text{Exp}\left[-\left(z^2+r_\text{FO}^2\right)\right] \left(\frac{r+\sqrt{r^2+z^2}}{r_\text{FO}+\sqrt{r_\text{FO}^2+z^2}}\right)^{C_\Gamma^{N_2} z^2/2}\,\nonumber \\ &\text{Exp}\left[- C_\Gamma^{N_2}/2\, \left(r\sqrt{z^2+r^2}-r_\text{FO} \sqrt{z^2+r_\text{FO}^2}\right)\right]\,,
\label{eq:fN2}
\end{align}
where a Maxwell-Boltzmann distribution for $N_2$ is assumed.\\

\begin{figure}[t]
  \includegraphics[width=0.48\textwidth]{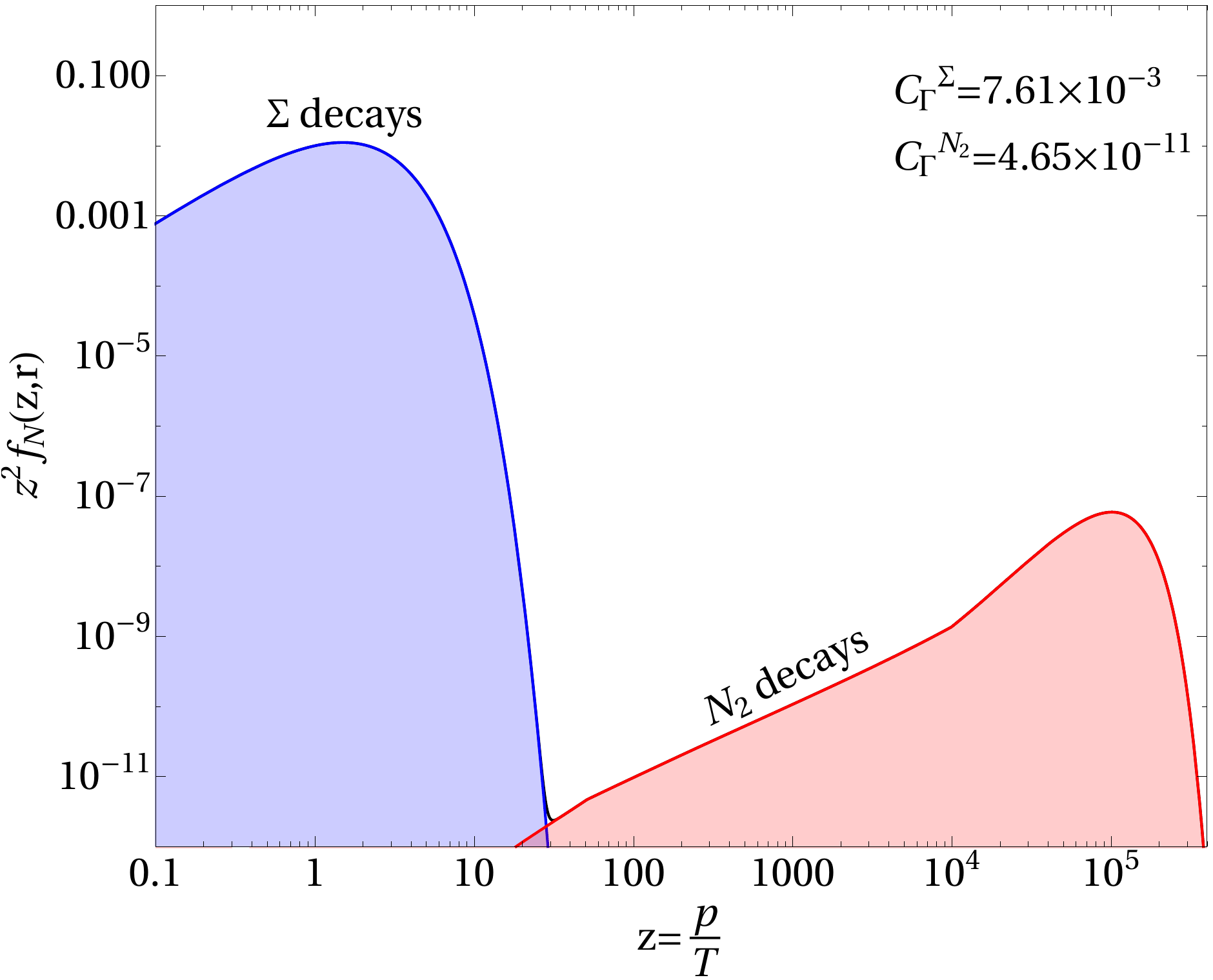}
  \includegraphics[width=0.48\textwidth]{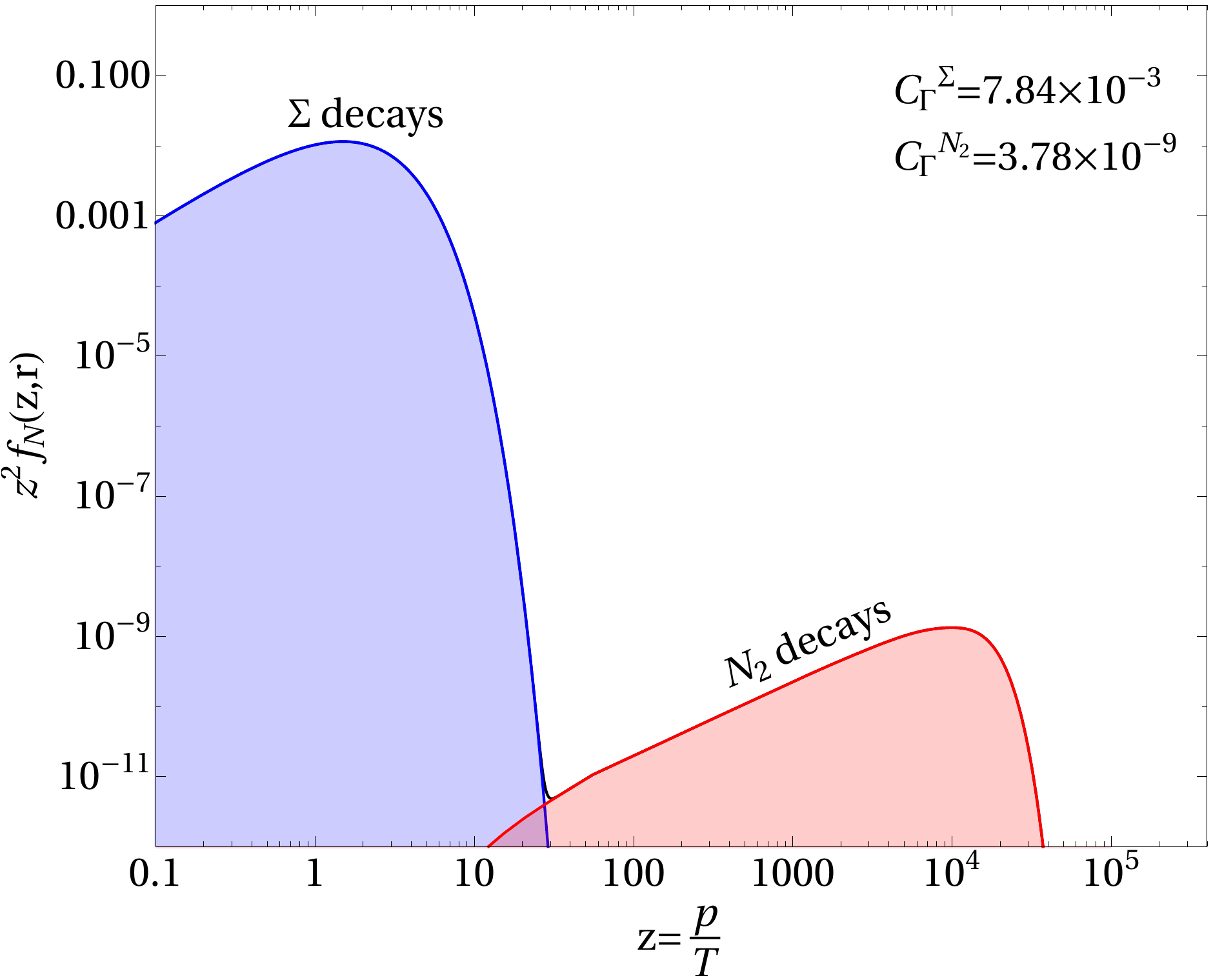}
  \caption{In the left (right) panel we show the momentum distribution function $f_{N_1}(z,r) \,z^2$ for both  DM production mechanisms, taking $m_\pm=600\GeV$ and $m_{N_2}=100\GeV\,(m_{N_2}=400\GeV)$. Blue and red curves correspond to freeze-in and $N_2$ late-time decays, respectively.}
  \label{fig:structure_form}
\end{figure}

In \cref{fig:structure_form} we show $f_{N_1}(z,r)\, z^2$ as a function of $z$ for two masses of $N_2$, namely $m_{N_2} =100\GeV$ and $400\GeV$, with scalar mass set to $m_\pm = 600\GeV$. The red curves represent $f^{N_2}_{N_1}(z,r)\,z^2$, obtained by solving \cref{eq:master} and fixing $r$ to sufficiently large values in order to capture the effect of decaying $N_2$. For comparison, we also show the distribution function corresponding to the production via freeze-in (blue), taking $r\to \infty$. Clearly, the peak of $f^{N_2}_{N_1}(z,r)\,z^2$ is shifted to very large values of $z$ indicating that $N_2$ decays yield a hot DM component. However, we also see from the figure that its amplitude is greatly suppressed with respect to $f^{\Sigma}_{N_1}(z,r)\,z^2$, implying that this component is subdominant for the selected benchmark point.
Quantitatively, the distribution shown in the left panel yields $ \Omega h^2_{N_2 \to N_1} = 0.03\times 0.12$, whereas for $m_{N_2}=400\GeV$ it follows that less than 1 per mille of the observed DM abundance is produced in $N_2$ decays. 

The simplest method for inferring the structure formation limit is to calculate free-streaming length by employing $\langle z \rangle$.
 However, this method is not applicable for our scenario since the spectrum consists of two components with two distinct peaks. 
Therefore we apply a more robust analysis and evaluate the transfer function $T(k)$, given by the power spectrum ratio

\begin{align}
    T^2(k) = \frac{P(k)}{P(k)_{\Lambda\text{CDM}}}\,,
\end{align}
where $P(k)$ is the power spectrum calculated from $f_{N_1}(z,r)$ using \texttt{CLASS} \cite{Lesgourgues:2011rh,Lesgourgues:2011re} and $P(k)_{\Lambda\text{CDM}}$ is the spectrum for cold DM only. The transfer function indicates at which scales non-cold DM will lead to deviations in comparison to cosmological observations.

The temperature of the DM species, $T_{N_1}$, relative to the photon temperature, $T_\gamma$, is relevant for the analysis and an input for \texttt{CLASS}: since we have two different mechanisms for DM production in the model, we are left with two independent dark sector temperatures:\\ 

$(i)$ One of these temperatures is set by the time when $N_1$ is produced via freeze-in mechanism from the decays of heavy scalars. These processes occur when the heavy scalars are still in thermal equilibrium implying that DM particles are produced with temperatures identical to those of SM sector. After production, $N_1$ is decoupled and does not experience reheating when SM degrees of freedom drop out of equilibrium. The temperature ratio, governed by the entropy dilution factor, yields
	\begin{align}
	T_{N_1}^{\text{freeze-in}} \approx \left(\frac{g_*(T_\nu)}{g_*(T^\text{prod})}\right)^{1/3}T_\gamma \approx \left(\frac{10.75}{114.25}\right)^{1/3} T_\gamma = 0.45\,T_\gamma\,,
	\end{align}
where $T^\text{prod}$ roughly corresponds to $\mathbf{Z}_2$-odd scalar masses.
	
$(ii)$ To evaluate the temperature of DM produced from out-of-equilibrium $N_2$ decays we estimate the temperature when these decays are taking place. We assume an instantaneous decay at $\tau = 1/\Gamma$ and make use of the time-temperature relation for a radiation-dominated Universe
	\begin{align}
	t = 2.42 \,\frac{1}{\sqrt{g_*(T)}} \left(\frac{1\,\text{MeV}}{T}\right)^2\, \text{s}\,,
	\end{align}
which allows us to obtain the expression for the temperature at which $N_2$ particles decay
\begin{align}
T_\Gamma = (g_*)^{-1/4} \left(\frac{\Gamma}{2.72\times 10^{-25}\GeV}\right)^{1/2}\,\text{MeV}\,.
\end{align}

For the benchmark point, employed already for presenting momentum distributions in the right panel of \cref{fig:structure_form}, we obtain
	\begin{align}
	\Gamma =  1.52\times 10^{-22}\GeV,\qquad \Omega h^2_{N_2}=5.61\times 10^3,\qquad T_\Gamma = 13\MeV.
\label{eq:values}
	\end{align}

At $T\sim T_\Gamma$, HNLs are at rest and each decay product has an energy $E\approx m_{N_2}/3\approx \mathcal{O}(100)\GeV$. Note that by dividing this energy with $T_\Gamma$ in \cref{eq:values} one obtains  
$z \simeq 10^{4}$ and this explains the position of the $N_2$ decay peak in the momentum distribution (see \cref{fig:structure_form}). In principle, the DM temperature is incorporated in the momentum distribution. However, we still have to take into account that, unlike DM, SM bath is reheated when electron-positron annihilation occurs. In summary, the temperature of this DM contribution is given by

	\begin{align}
	T_{N_1}^\text{decay} \simeq \frac{m_{N_2}}{3T_\Gamma} \left(\frac{4}{11}\right)^{1/3} \, T_\gamma \approx 7300\, T_\gamma\,,
	\end{align}
where the expression is evaluated for the aforementioned benchmark point.\\

\begin{figure}[t]
  \includegraphics[width=0.65\textwidth]{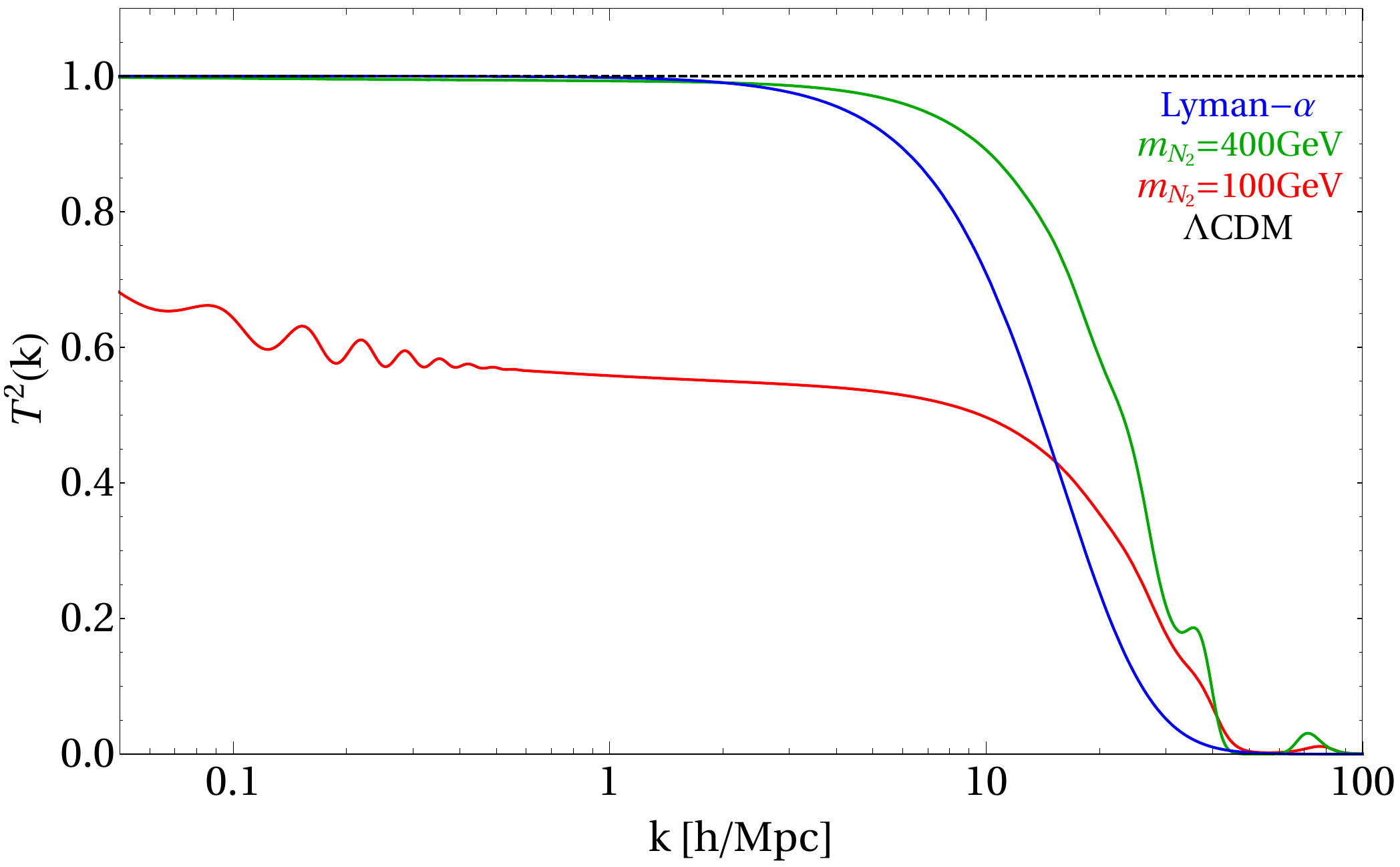}
  \caption{Transfer function for the same benchmark points as in \cref{fig:structure_form}. The constraint from structure formation, using $m_\text{TR}=2\keV$ is shown in blue. HNL masses of around $100$ GeV clearly violate this constraint while the green line (corresponding to $m_{N_2}=400\GeV$) is consistent with the data. The black dashed line represents $\Lambda$CDM.}

  \label{fig:transfer_function}
\end{figure}

In order to assess the cosmological viability of particular benchmark points, we compare the calculated $T^2(k)$ against the function corresponding to the constraint stemming from Lyman-$\alpha$ forests. For the latter, we adopt an analytical fit for the transfer function \cite{Viel:2005qj}, taking $m_\text{NRP}\simeq 10\keV$\footnote{We fixed the mass of thermal relic $m_\text{TR}=2$ keV, and the relation with the mass of non-resonantly produced particle is given by $m_\text{NRP} = 4.35 \, (m_\text{TR}/\keV)^{4/3}$ \cite{Heeck:2017xbu}.}. In \cref{fig:transfer_function} we show in red (green) the calculated transfer function for $f_{N_1}(z,r)$ with $m_\pm = 600\GeV$ and  $m_{N_2}=100\GeV\,(m_{N_2}=400\GeV)$; these are identical benchmark points as those from \cref{fig:structure_form}. If a given curve lies below the Lyman-$\alpha$ limit (blue curve) the corresponding parameter point is disfavored. We observe that the scenario with lighter mass of $N_2$ is excluded since the abundance of hot DM is too large in this case and hence larger cosmological scales than observed are affected. On the other hand, the green curve is in agreement with observational data. One should note that curves drop to zero at roughly the same point; this scale is set by the temperature of the dominant, frozen-in, DM component. If $N_1$ would freeze-in at later times the curves would shift to the left.\\

\subsection{Constraints from $N_\text{eff}$}
\label{subsec:N_eff}
\noindent
When discussing possible implications on the formation of structures in the early Universe, one should also take into account that keV-scale DM could change the number of relativistic species $N_\text{eff}$. 
The effective number of relativistic non-photonic species, $N_\text{eff}$, after electron-positron annihilation, enters in the expression for the radiation density
	\begin{align}
	\rho_\text{rad}= \left[ 1+ \frac{7}{8}\left(\frac{4}{11}\right)^{4/3}  N_\text{eff}\right] \, \rho_\gamma\,,
	\end{align}
where $\rho_\gamma$ represents the energy density of photons. In the SM, $N_\text{eff} = 3.046$ and thus we denote contributions from additional relativistic species as $\displaystyle \Delta N_\text{eff} = N_\text{eff}-3.046$. The contribution to $\Delta N_\text{eff}$ from $N_1$ can be estimated by comparing its energy density against the one corresponding to a fully relativistic neutrino with temperature $T_\nu$
 \cite{Merle:2015oja}:
	\begin{align}
	\Delta N_\text{eff}(T_\nu) = \frac{60}{7\pi^4} \frac{m_{N_1}}{T_\nu}\int\limits_{0}^{\infty} \left[  \sqrt{1 + \left( \frac{z\,T_\nu}{m_{N_1}}\right)^2}-1\right]\,z^2f_{N_1}(z,T_\nu)\,\dd z \,\times\begin{cases}
	1,&	\text{if}\; T_\nu>1\MeV\\
	\left(\frac{11}{4}\right)^{4/3},&	\text{if}\; T_\nu<1\MeV
	\end{cases}.
	\label{eq:delta_neff}
	\end{align}
	\begin{figure}[t]
		\includegraphics[width=0.5\textwidth]{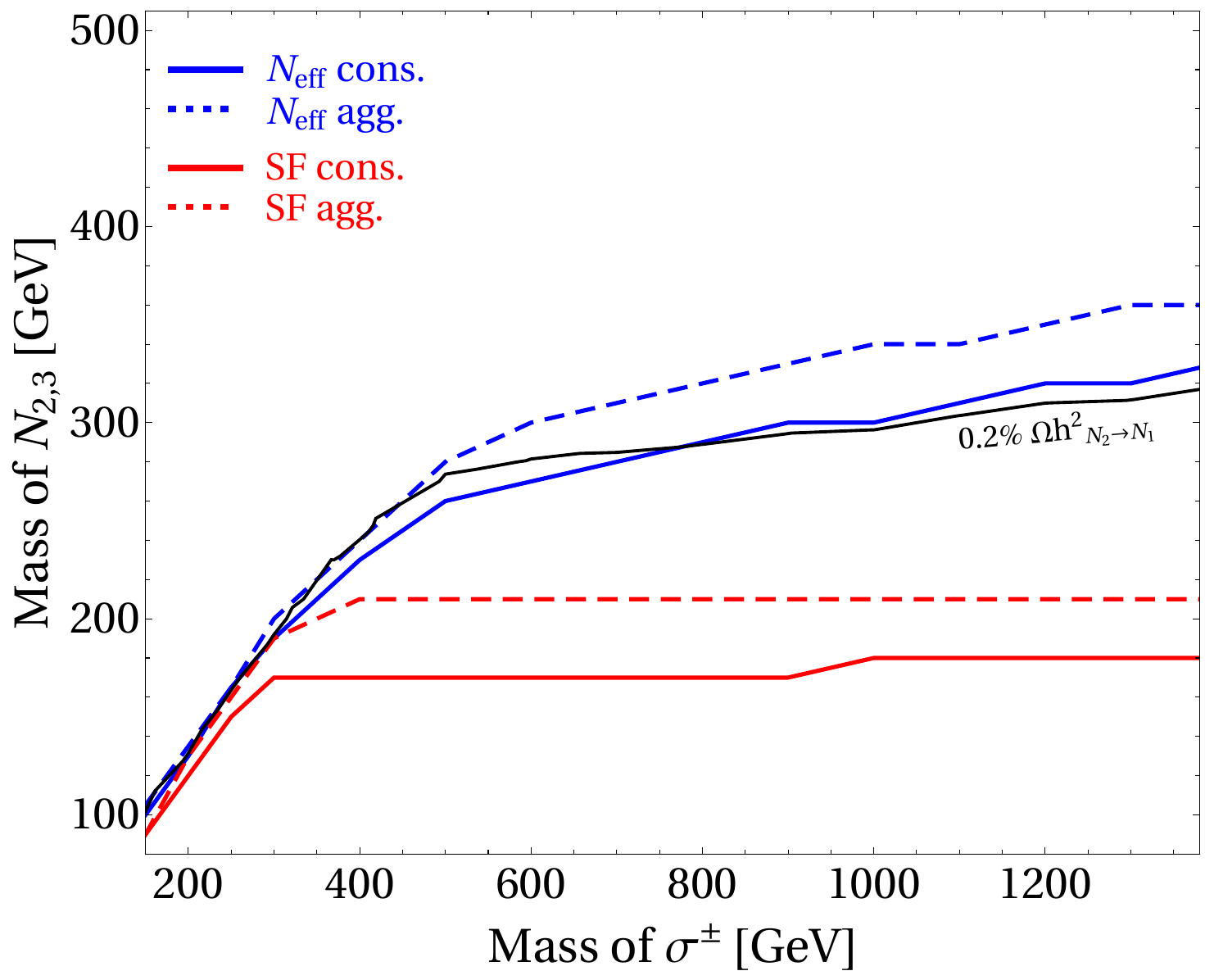}
		\caption{Constraints from structure formation (red curves) confronted with $N_\text{eff}$ limits (blue curves) derived using \cref{eq:delta_neff}. The solid curves correspond to the conservative and dashed ones to the aggressive choice of $\Delta N_\text{eff}$. Note that structure formation limits also indirectly depend on $\Delta N_\text{eff}$ as it is an input parameter for \texttt{CLASS}. Clearly, $N_\text{eff}$ yields much stronger limits in comparison to those arising from structure formation. Shown in black is the curve for $\Omega h^2_{N_2\to N_1}=0.24\times 10^{-3}$.}
		\label{fig:SF_Neff_comb}
	\end{figure}
	
Using as an example the benchmark points employed in \cref{fig:structure_form}, we can derive the following values:
\begin{align}
	& m_{N_2}=100\GeV \rightarrow \Delta N_\text{eff}\sim 220\,,  &
	m_{N_2}=400\GeV\rightarrow  \Delta N_\text{eff}\sim0.02\,.
	\label{eq:Neff}
\end{align}

Clearly, large mass gaps between $N_2$ and $\sigma^\pm$ are disfavored; the reason is that such cases would lead to larger abundances of $N_2$ and therefore the hot DM component becomes more prominent in the spectrum.

Current measurements by the Planck collaboration allow for an upper limit of $\Delta N_\text{eff} \approx 0.28$ at 95\% CL (TT, TE, EE+lowE+lensing+BAO). Including the present tension in the Hubble constant measurement, this value  increases to $\Delta N_\text{eff} \approx 0.52$ at 95\,\% CL (TT, TE, EE+lowE+lensing+BAO+R18) \cite{Aghanim:2018eyx}. In the following, we dub the first value \textit{aggressive} and the second one \textit{conservative}.

By using \cref{eq:delta_neff}, we can estimate which parameter choices lead to $\Delta N_\text{eff}$ values that exceed the conservative/aggressive value.
We have performed a scan and have deduced the following conditions when using conservative values 
\begin{align}
    \frac{\Omega h^2 _{N_2\to N_1}}{\Omega h^2_\text{DM}} \lesssim 0.2\%\,, \quad C_\Gamma^{N_2} \gtrsim  5\times10^{-10}\GeV\,,
\end{align}
necessary for consistency with cosmology (see also black curve in \cref{fig:SF_Neff_comb}). 
For instance, taking $m_\pm = 1\TeV$, the lower bound on the heavy lepton mass is $m_{N_2}\gtrsim 300\GeV$\footnote{There is a caveat as one has a freedom to choose the couplings between $N_{2,3}$ and the charged leptons. Throughout this section we assume couplings to $\tau$ leptons to be subdominant.}. 

In \cref{fig:transfer_function} and \cref{eq:Neff} we demonstrated that one of the chosen benchmark points is excluded by both structure formation and $N_\text{eff}$ considerations.
Comparing both probes in \cref{fig:SF_Neff_comb} we however conclude that $N_\text{eff}$ generally leads to stronger exclusion limits. Hence, in \cref{sec:pp-collider,sec:future} we will compare regions in parameter space that are accessible at colliders with $ N_\text{eff}$ limits.


\subsection{BBN constraints}
\label{subsec:BBN}
\noindent
Primordial abundances of light nuclei may be affected by processes involving new particles in the model. As we have seen in the previous section, $N_2$ decays produce SM leptons with large momenta, which can inject a lot of energy into the plasma. Specifically, we need to ensure that $N_2$ decays are fast enough such that these highly energetic particles can thermalize with the plasma and thus the BBN measurements remain unaffected. $N_2$ decays to $N_1$ and a pair of leptons and the rate of this process is given in \cref{eq:decay_N2}, being proportional to two powers of small Yukawa coupling $y_{1\alpha}$. 

In order to obtain the BBN limits in the considered scenario, we adopt the results from the recent analysis \cite{Kawasaki:2017bqm} where the authors studied the impact of the decaying hidden sector particles to the primordial abundances of light nuclei. The channels of our interest are those containing charged leptons.
Decays of $N_2$ into electrons and muons can take as long as $\mathcal{O}(100\,\text{s})$, since they mainly induce electromagnetic showers, which affect BBN at later times only. In contrary, $\tau$ leptons decay mostly hadronically and this can significantly alter the observed neutron to proton ratio; unless the abundance of $N_2$ is strongly suppressed, $N_2$ decay time has to be $\lesssim 1\,\text{s}$.

\begin{figure}[t]
	\centering
	\includegraphics[width=0.48\textwidth]{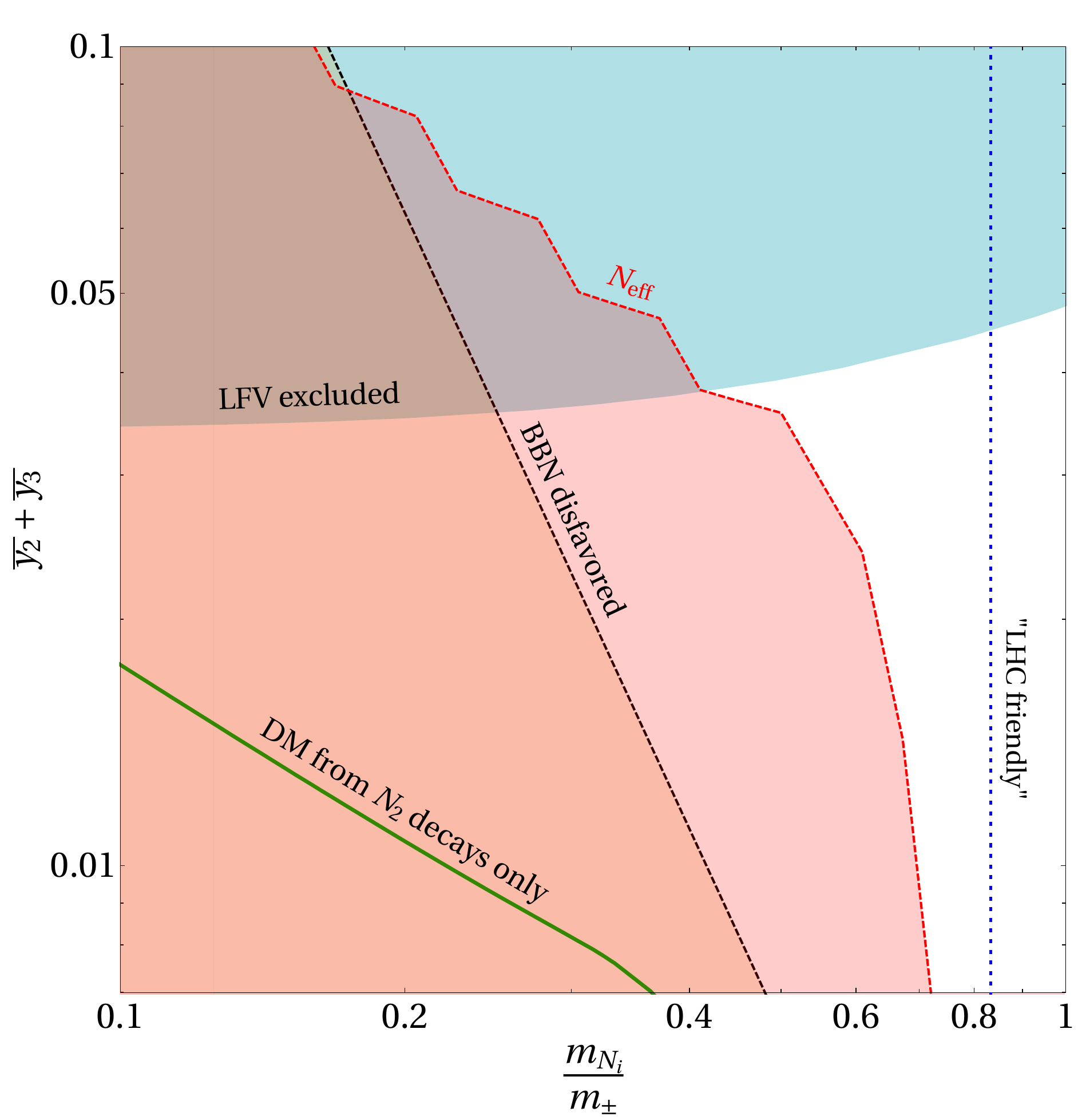}
	\includegraphics[width=0.48\textwidth]{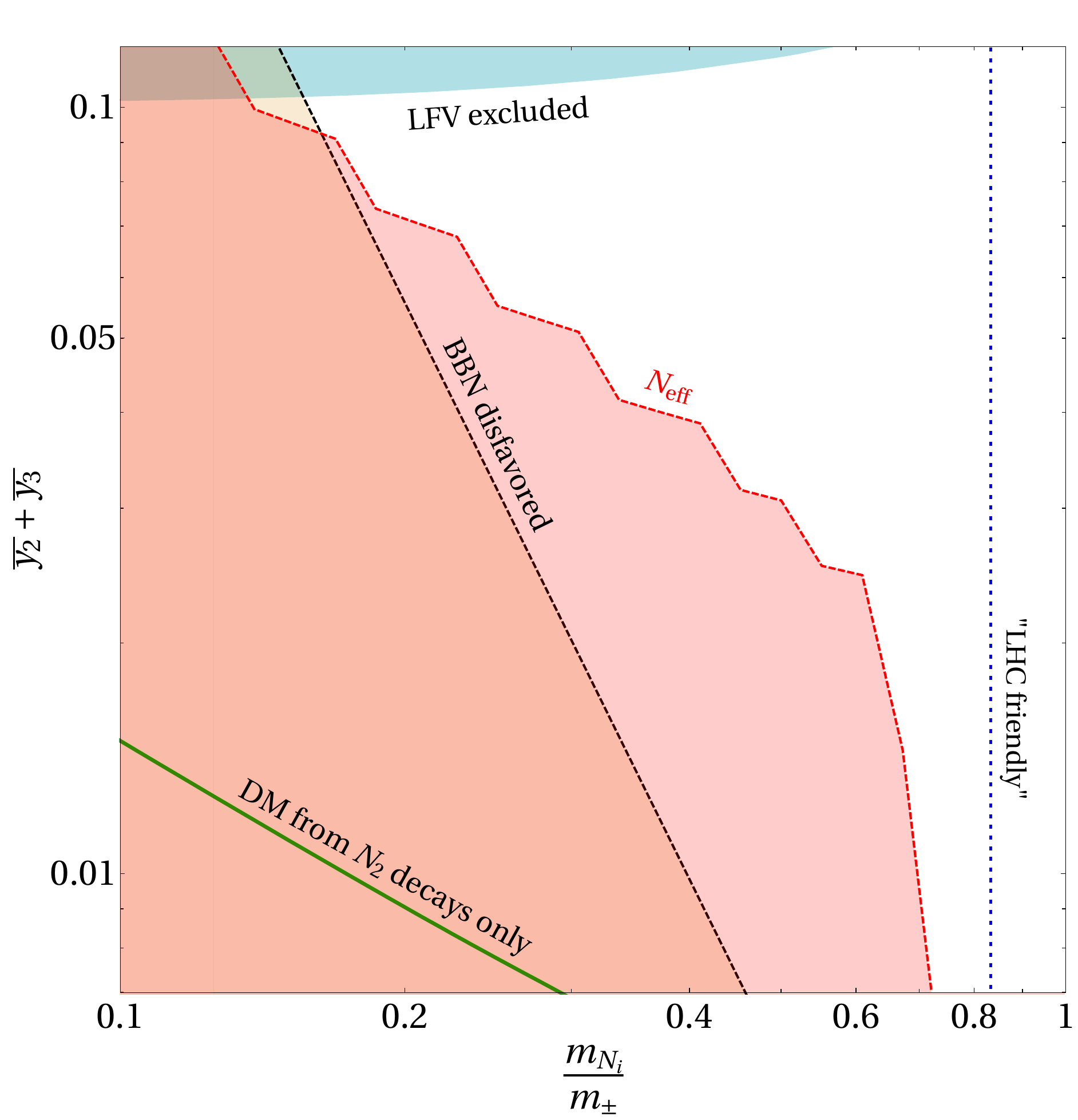}
	\caption{BBN constraints for the case where $N_2$ dominantly decays into electrons and muons (left panel) and tau-leptons (right panel) are disfavoring parameter space below the black lines. The regions excluded by a conservative $N_\text{eff}$ limit are shown in red. The solid green curve indicates the parameter space for a limiting case in which all of the DM is produced by $N_2$ decays. The regions in blue represent constraints from LFV experiments. 
Finally, the region to the left of the vertical blue dashed line is collider friendly in the sense that 
the mass gap between $\sigma^\pm$ and $N_{2,3}$ is sufficiently large. The value of the charged scalar mass in both panels is fixed to $m_\pm=600\GeV$. On the y-axis we show the average Yukawa coupling of $N_2$ and $N_3$, defined as $\bar{y}_2+\bar{y}_3 \equiv \sqrt{(1/3)\sum_\alpha \left(|y_{2\alpha}|^2 +|y_{3\alpha}|^2\right)}$.}
	\label{fig:BBN}
\end{figure}
In \cref{fig:BBN}, black lines indicate BBN exclusions for a representative value of $m_\pm=600$ GeV. The left panel corresponds to the dominant decay into $e^\pm/\mu^\pm$ and in the right panel the case where $N_2$ decays prominently into $\tau^\pm$ pairs is shown. These channels are motivated by the di-lepton and di-tau searches at colliders which will be presented in the following sections. 
To be conservative, we impose the decay time for the respective channel to be shorter than $1$ second. 

Note that by increasing the scalar masses, larger $y_{1\alpha}$ are required for DM production through $\Sigma$ decays and hence the BBN bounds get weaker. LFV bounds (blue regions) are then also relaxed, see \cref{eq:LFV}.

The thick green solid line indicates a parameter space corresponding to the DM production only through $N_2$ decays and such scenario is clearly excluded; this curve lies below the red shaded region which indicates $N_\text{eff}$ limits. We found numerically that in terms of DM density, the $N_\text{eff}$ exclusion line is roughly set by the requirement $\Omega h^2_{N_2\to N_1}/\Omega h^2_\text{DM}\approx0.1\%$. The blue vertical dashed line shows the region where the energies of final state leptons, arising from $\sigma^\pm \to N_2\, l_\alpha$ are below $100\GeV$, making them harder to resolve at colliders.

Here we have shown that there are regions which are not excluded neither by LFV nor $N_\text{eff}$ and BBN arguments. In the next two sections we will demonstrate that some of these portions in the parameter space overlap with the regions accessible at colliders.

\section{HL-LHC Projections}
\label{sec:pp-collider}
\noindent
In this section we explore the di-lepton and di-tau signatures with missing transverse energy:

\begin{align}
 p \,p \to\sigma^+  \sigma^-\to 
\begin{cases}
    \sigma^\pm \to l^\pm \, N_i \, (l=e^\pm \,\text{or} \, \mu^\pm)\,,\\
    \sigma^\pm \to \tau^\pm \, N_i  \,.
\end{cases}
\end{align}

 The former were for instance already scrutinized for the scotogenic model in \cite{Hessler:2016kwm}, where the authors used the data from LHC Run-1 to set the limits. Here we

\begin{itemize}
\item calculate the projected sensitivities for HL-LHC using the same analysis techniques as presented in recent ATLAS publications (Ref.~\cite{Aaboud:2018jiw} and Ref.~\cite{Aaboud:2017nhr} for di-lepton and di-tau channel, respectively). In particular, we are using the results from Run-2 with an integrated luminosity of $36.1\invfb$ at $\sqrt{s}=13\TeV$.

\item present both \emph{optimal} and \emph{realistic} projections; the first one is defined such that the branching ratio for charged scalar decay into HNL and charged lepton considered in the search is set to $1$. Such case is, however, not feasible in our model\footnote{We generate Yukawa matrices using Casas-Ibarra parametrization (\cref{sec:model}) in order to obtain viable values for neutrino masses and mixings.} and hence we also define realistic projections by maximizing respective Yukawa couplings. This will lead to branching ratios smaller than 1.

\item are in position to confront the calculated projections with bounds from BBN and structure formation (see \cref{sec:DM}). Let us point out that this is very rarely performed in the papers focused on collider signatures; in particular this \emph{has not} been done for the scotogenic model.

\end{itemize}

\begin{figure}[ht!]
   \centering
	\includegraphics[width=.45\textwidth]{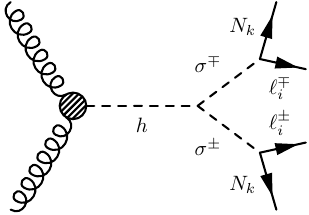}
	\includegraphics[width=.45\textwidth]{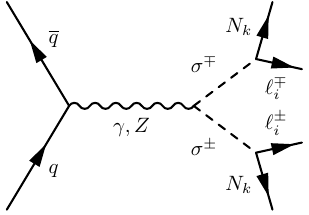}
  \caption{Production channels for the $\ell_i\,\ell_j\mathbf{+\cancel{E_T}}$ process at the LHC. Pair produced charged scalars decay into heavy leptons $N_{2,3}$ and SM charged leptons $(e,\mu$ or $\tau)$.}
  \label{fig:Production_process}
\end{figure}
The model files were created with \texttt{FeynRules} \cite{Alloul:2013bka}. The signal processes, as shown in \cref{fig:Production_process}, were simulated at leading order (LO) with \texttt{MadGraph5\_aMC@NLO\_2.6.3.2} \cite{Alwall:2014hca} interfaced with \texttt{Pythia8} \cite{Sjostrand:2007gs} for showering the events and \texttt{Delphes\_3.4.1} \cite{deFavereau:2013fsa} was used for a fast detector simulation. By using these tools, we were able to reproduce the results given in \cite{Aaboud:2017nhr,Aaboud:2018jiw}. 

\begin{figure}[ht!]
	\centering
	\includegraphics[width=.49\textwidth]{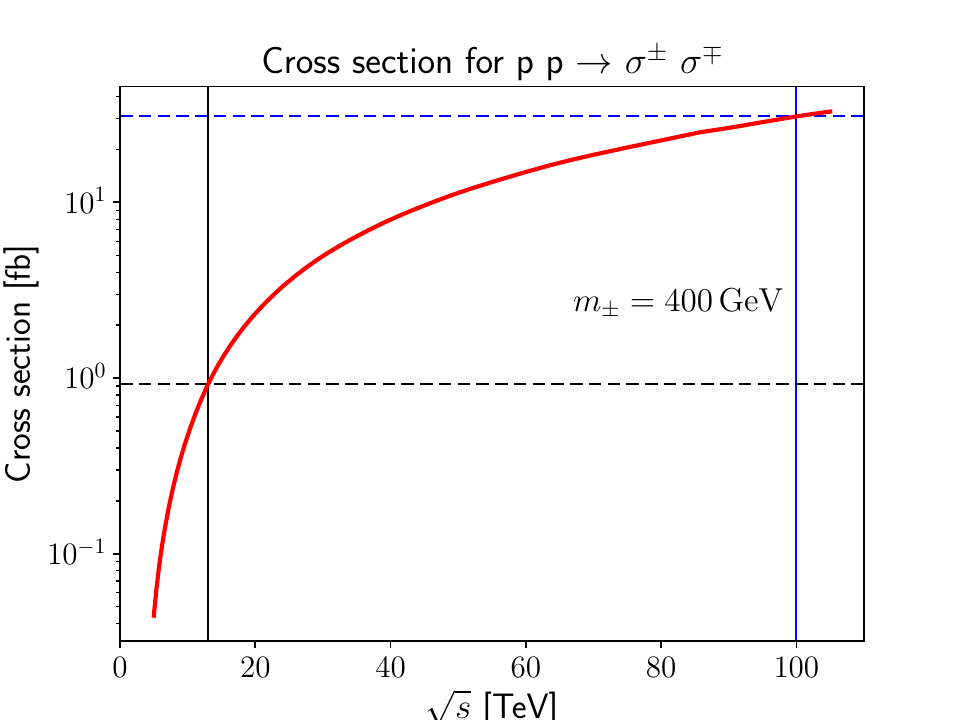}
	\includegraphics[width=.49\textwidth]{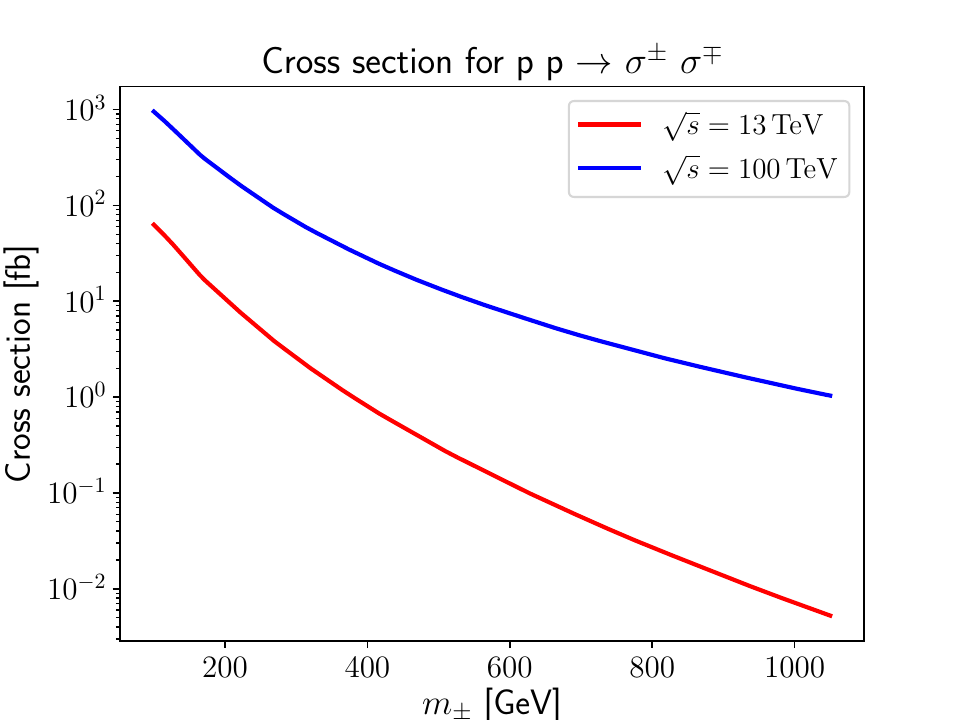}
	\caption{Cross section $\sigma^\text{prod}$ for pair production of charged scalars $\sigma^\pm$. The left panel shows the increase of $\sigma^\text{prod}$ for larger center of mass energies and fixed scalar mass. The black (blue) vertical lines indicate the energy range of HL-LHC (FCC-hh) while dashed lines correspond to the respective cross section. In the right panel we show $\sigma^\text{prod}$ for different scalar masses and fixed energy. By increasing $m_\pm$ the cross section drops significantly.}
	\label{fig:pair_prod_xsec}
\end{figure}

In \cref{fig:pair_prod_xsec} we show the expected cross section for $p\,p\to \sigma^\pm\,\sigma^\mp$ pair production for different energies and masses. One can already conclude that large luminosities are needed to see a significant number of events inside the detector. For definiteness we have fixed the portal couplings to 
\begin{equation}
\lambda_3 = 0.3\,,\quad \lambda_4 = 0.5\,,\quad \lambda_5 = 10^{-4}\,.
\end{equation}
The most relevant parameter in the scalar sector is the physical mass of the charged scalar, $m_\pm$, and it is this quantity that will appear in all our sensitivity projections. 

We assume that $m_{N_2}=m_{N_3}$ but in general one can take a hierarchical spectrum as well, \emph{i.e.} $m_{N_2}<m_{N_3}$. A hierarchical spectrum would, on the one hand, weaken the search strategy because in this case decays $\sigma^\pm \to \ell^\pm\,N_3$ are more likely to yield soft leptons due to the smaller mass gap between $\sigma^\pm$ and $N_3$. On the other hand, this could give rise to interesting event topologies, because $N_3$ can decay in the detector into 2 leptons and $N_2$ and this would possibly create multi-lepton + $\cancel{E_T}$ signatures. We leave the study of such scenario for future work and in what follows focus on the di-tau\met and di-lepton\met searches with the degenerate spectrum.


\subsection{Di-tau$\mathbf{+\cancel{E_T}}$}
\label{subsec:taus}
\noindent
Following the procedure outlined in \cite{Aaboud:2017nhr}, the following cuts were applied after event reconstruction: Events shall contain no b-jet but at least two tau leptons with opposite charges. The invariant mass of every tau pair has to be larger than $12\GeV$ and must be $10\GeV$ away from the mean visible Z boson mass, set at $79\GeV$. Then two different trigger setups are defined. The asymmetric trigger requires $p_T^1 > 85\GeV$ and $p_T^2>50\GeV$ for the first two $p_T$ ordered $\tau$ leptons. Second, there is the $\cancel{E_T}$ trigger set by $p_T^1>35\GeV$, $p_T^2>25\GeV$ and $\cancel{E_T}>50\GeV$. For further discrimination from SM background, the stransverse mass \cite{Barr:2003rg}, $m_{T_2}$,  is introduced.\\
Finally, two signal regions based on $m_{T_2}$ and $\cancel{E_T}$ cuts were defined as shown in \cref{tab:SR_Ditau_definition}.
\begin{table}[t]
\centering
\begin{tabular}{|c|c|c|}
\hline
    SR-lowMass & \multicolumn{2}{c|}{SR-highMass}  \\
    \hline
    $m_{T_2}>70\GeV$ & \multicolumn{2}{c|}{$m_{T_2}>70\GeV$ \& $m(\tau_1,\tau_2)>110\GeV$}\\
      $\cancel{E_T}$ trigger  &   $\cancel{E_T}$ trigger  &   asymmetric trigger \\
    $\cancel{E_T}>150\GeV$  &   $\cancel{E_T}>150\GeV$  &   $\cancel{E_T}>110\GeV$\\
    $p_T^1>50\GeV$          &   $p_T^1>80\GeV$          &   $p_T^1>95\GeV$\\
    $p_T^2>40\GeV$          &   $p_T^2>40\GeV$          &   $p_T^2>65\GeV$\\
    \hline
\end{tabular}
\caption{Signal regions used in the di-tau analysis.}
\label{tab:SR_Ditau_definition}
\end{table}
They are chosen such that different mass gaps between $\sigma^\pm$ and $N_{2,3}$ can be covered. The
95\% CL upper limits on the cross sections are summarized in \cref{tab:SR_Ditau}.
\begin{table}[t]
 \centering
 \begin{tabular}{|c|c|c|c|c|}
 \hline
  signal region	& $N_\mathrm{exp}$	& $N_\mathrm{obs}$	& exp. $\sigma^{95}_\mathrm{vis}$\,[fb]	& obs. $\sigma^{95}_\mathrm{vis}$\,[fb]\\
  \hline
  SR-lowMass	& 	$14\pm 6$	&	10		&	$0.31^{+0.12}_{-0.08}$		& 0.26\\
  SR-highMass	& 	$3.7\pm1.4$	&	5		&	$0.17^{+0.08}_{-0.05}$		& 0.20\\
  \hline
 \end{tabular}
  \caption{95\% CL limits on the non-SM cross section for the di-tau \met analysis.}
  \label{tab:SR_Ditau}
\end{table}\newline
The final cross section in our model is determined by the product
  \begin{equation}
   \sigma( p\,p \to \ell^\pm \ell^\mp N_k\,N_l) = \sigma( p\,p \to \sigma^\pm \sigma^\mp) \times \mathrm{BR}(\sigma^\pm \to \ell^\pm_i N_k)^2. 
  \end{equation}
  The cross section can be increased by maximizing the respective branching ratios. The latter can be achieved by making use of unconstrained parameters. In particular, we  took the complex angle $\theta=\omega-i\eta$ (see \cref{eq:CIR}), the Majorana phase $\alpha_2$\footnote{The other Majorana phase does not affect the minimization.} and the CP phase $\delta$\footnote{We allowed $\delta$ to float in the range $(135^\circ, 366^\circ)$ which corresponds to a  $3\sigma$ range from recent fits.} as free parameters
 and maximized the following expression
  \begin{equation}
   \frac{y_{2\tau}(\omega,\eta,\alpha_2,\delta)^2+y_{3\tau}(\omega,\eta,\alpha_2,\delta)^2}{\sum\limits_{k=2,3}\sum\limits_i y_{ki}(\omega,\eta,\alpha_2,\delta)^2}\,.
   \label{eq:optimization_eq}
  \end{equation}
The parameter values that correspond to the extremum are in what follows denoted as $\omega_0,\eta_0,\alpha_2^0,\delta_0$. We have performed this procedure for both normal and inverted neutrino mass ordering. The results are given in \cref{tab:BR_Ditau}.\\
\begin{table}[t]
 \centering
 \begin{tabular}{|c|c|c|c|c|c|}
 \hline
	&	$\omega_0$	&	$\eta_0$		&	$\alpha_2^0$	&	$\delta_0$	&	$\BR(\sigma^\pm \to \tau^\pm N_k)$\\
    \hline
    NO	&	1.61	&	$>2$		&	$\pi$		&	$2\pi$		&	38.26\,\%\\
    IO	&	1.31	&	$>2$		&	$-\pi$		&	$\pi$		&	27.30\,\%\\\hline
 \end{tabular}
  \caption{Largest possible branching ratios for the decay of $\sigma^\pm$ into $\tau$ and $N_{2,3}$. Above the given value for $\eta_0$, the branching ratios are to a good approximation independent of this parameter. As can be seen, the IO  gives rise to smaller branching ratios compared to NO.}
  \label{tab:BR_Ditau}
\end{table}

For calculating the sensitivity curves we fixed the portal couplings $\lambda_i$ such that $\sigma^\pm$ is the lightest $\mathbf{Z}_2$-odd scalar, suppressing the additional decays into the other scalars. We scanned over the charged scalar mass as well as the heavy lepton masses $m_{N_{2,3}}$; our grid spans $m_\pm \in (150\GeV,\,600\GeV)$ and $m_{N_{2,3}} \in (10\GeV,\, m_\pm)$ and we simulated $10^4$ events for each point.

We compared the simulation with the recent ATLAS results and found that current sensitivities are not strong enough to place limits. The reason is twofold: First, the cross section for the pair production in the model is significantly smaller than the one in the simplified model used in the ATLAS analysis. Second, the analysis uses specific cuts on kinematic
variables which do suppress SM background but  unfortunately also cut away a significant portion of signal events. For instance, the ``best case" benchmark point (where we set the branching ratio into tau leptons to 1) features a quite small surviving cross section:
\begin{align}
 \mathrm{Benchmark}:\quad m_{N_{2,3}}=10\GeV\quad m_\pm = 200\GeV \rightarrow \text{SR-highMass:\;}\sigma_\mathrm{vis}=(0.15\pm 0.06)\fb.
\end{align}
After taking Casas-Ibarra parametrization into account, and inserting realistic branching ratios, the situation gets even worse as the cross section is further reduced due to non-maximal branching ratio into tau leptons. This finding motivated us to
go beyond the current experimental results and consider a similar search at the foreseen HL-LHC program \cite{Apollinari:2017cqg} at CERN which will deliver a final integrated luminosity of up to $4000\invfb$ at $\sqrt{s}=14\TeV$. This would lead to a huge increase in potential signal events.

To estimate the potential of HL-LHC to test the scotogenic model we conduct a similar analysis as in \cite{Aaboud:2017nhr} but use a projected sensitivity
\begin{align}
\mathcal{S} = \frac{S}{\sqrt{S+B}}\,, 
\label{eq:ssb}
\end{align}
where $S$ and $B$ represent signal and background events, respectively. This formula is derived in the limit $S/B\ll 1$ from the general expression for the case of exclusion limits 
\begin{align}
\mathcal S_1=\sqrt{2\left(S-B\log\left(1+\frac SB\right)\right)}\,, \label{eq:S1}
\end{align}
obtained using the procedure described in \cite{Cowan:2010js}.

\begin{figure}[t]
 \centering
 \includegraphics[width=0.5\textwidth]{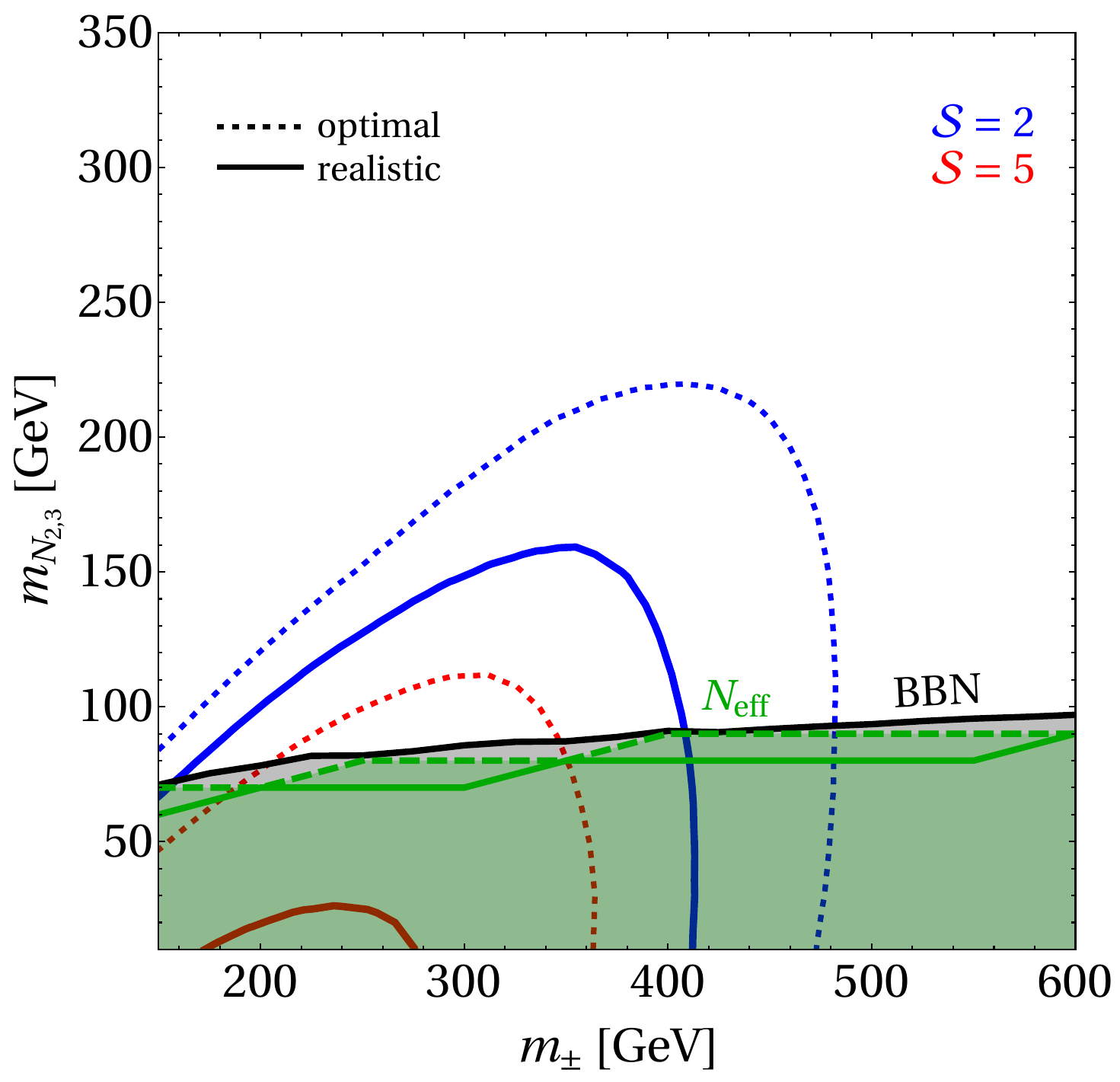}
 \caption{Projected sensitivities for HL-LHC using $\mathcal{L}=4000\invfb$ and the same analysis techniques as in Ref.~\cite{Aaboud:2017nhr}. Given in blue are exclusion limits, $\mathcal{S}=2$, and shown 
 in red are discovery limits, where $\mathcal{S}=5$. The dashed lines correspond to a 100\% branching ratio into tau leptons, whereas the solid lines represent the case in which the maximized branching ratio for NO (shown in \cref{tab:BR_Ditau}) is employed. HNL masses in the black shaded region are in conflict with BBN constraints discussed in \cref{subsec:BBN} and the green lines arise from $N_\text{eff}$ limits: the dashed green line indicates conservative and the solid one represents aggressive limit, as discussed in \cref{subsec:Lyalpha}.}
 \label{fig:Ditau_excl}
\end{figure}

By taking the identical signal regions as in the previous analysis and assuming a similar scaling of signal and background for the increased center of mass energies and luminosities we can now redo the cut and count analysis for increased event rates.  As can be seen in \cref{fig:Ditau_excl} (solid lines), this allows us to significantly enhance the testable parameter space. For such high luminosities, scalar masses
of up to 420\,GeV and respective HNL masses of 170\,GeV can be tested and there is even a potential discovery region for scalar masses between $200$ and $300$ GeV. The sharp drop for large masses is due to a decrease in the pair production cross section. The cuts on the kinematic variables, such as transverse momentum $p_T$ and stransverse mass $m_{T2}$, need a sufficiently large mass gap between $\sigma^\pm$ and $m_{N_{2,3}}$ which bounds the accessible parameter space from above and also from the left, because charged scalar should not be too light, as in this regions leptons are too soft. We also show in dashed corresponding 
sensitivity curves for the optimal case. As expected, the sensitivities improve in this case, however we have not found such scenario in our numerical procedure (see \cref{tab:BR_Ditau}); the branching ratios for the decay into pair of tau leptons can be at most around $40\%$. By taking BBN and $N_\text{eff}$ limits into account it turns out that the discovery region $(\mathcal{S}\geq 5)$ is in tension with these cosmological probes; only parameter space around $m_{N_{2,3}}\simeq 100$ GeV is not excluded for the optimal case while the sensitivity for the realistic case is strongly ruled out. However, a certain portion of parameter space with $\mathcal{S}\gtrsim 2$  around 
$m_{N_{2,3}}\simeq 100-150$\,GeV and $m_\pm \simeq$ $200-400$\,GeV (realistic case) is not constrained by cosmological data.

\subsection{Di-lepton$\mathbf{+\cancel{E_T}}$}
\label{subsec:leptons}
\noindent
Now we turn to di-lepton\met channel. Following the procedure outlined in Ref. \cite{Aaboud:2018jiw}, the following cuts were applied after event reconstruction and preselection. The invariant di-lepton mass should be larger than $40\GeV$. Events should not contain any b-jet with $p_T>20\GeV$ nor a jet with $p_T>60\GeV$. 
In the $2\ell + 0$jets channel, six different signal regions were defined: $4$ aiming at different flavor (DF) leptons in the final states and $2$ at leptons with the same flavor (SF). All regions are
inclusively defined and mainly separated by increasing cuts on the invariant mass of the lepton pair and $m_{T2}$, ranging from $m_{\ell \ell}>110\GeV$ to $m_{\ell \ell}>300\GeV$ and $m_{T2}>100\GeV$ to $m_{T2}>300\GeV$. The 95\% CL upper limits on the cross sections are summarized in \cref{tab:SR_Dilep}.\\
\begin{table}[t]
 \centering
 \begin{tabular}{|c|c|c|c|c|}
 \hline
  signal region	& $N_\mathrm{exp}$	& $N_\mathrm{obs}$	& exp. $\sigma^{95}_\mathrm{vis}$\,[fb]	& obs. $\sigma^{95}_\mathrm{vis}$\,[fb]\\
  \hline
  SF-loose	& 	$133\pm 22$	&	153		&	$1.47^{+0.58}_{-0.44}$		& 2.02\\
  SF-tight	& 	$9.8\pm2.9$	&	9		&	$0.33^{+0.11}_{-0.08}$		& 0.29\\
  DF-100 	&	$68\pm7$	&	78		&	$0.75^{+0.30}_{-0.22}$		& 0.88\\
  DF-150 	&	$11.5\pm3.1$	&	11		&	$0.33^{+0.14}_{-0.11}$		& 0.32\\
  DF-200 	&	$2.1\pm1.9$	&	6		&	$0.29^{+0.08}_{-0.05}$		& 0.33\\
  DF-300 	&	$0.6\pm0.6$	&	2		&	$0.16^{+0.03}_{-0.02}$		& 0.18\\
  \hline
 \end{tabular}
  \caption{95\% CL limits on the non-SM cross section for the di-lepton \met analysis.}
  \label{tab:SR_Dilep}
\end{table}

In contrast to \cref{subsec:taus} we now want to minimize the expression given in \cref{eq:optimization_eq} in order to have as large as possible branching ratios into leptons. Under the same assumptions as before and
taking both, NO and IO regimes into account we obtain the results given in \cref{tab:BR_Dilep}.
\begin{table}[t]
 \centering
 \begin{tabular}{|c|c|c|c|c|c|}
  \hline
	&	$\omega_0$	&	$\eta_0$		&	$\alpha_2^0$	&	$\delta_0$	&	$\BR(\sigma \to \ell_i N_k)$\\
    \hline
    NO	&	2.37	&	$<-2$		&	$\pi$		&	$2\pi$		&	86.42\,\%\\
    IO	&	3.07	&	$<-2$		&	$-\pi$		&	$\pi$		&	99.75\,\%\\ 
    \hline
 \end{tabular}
  \caption{Largest possible branching ratios for the decay of $\sigma^\pm$ into $e^\pm,\mu^\pm$. Below the given value for $\eta_0$, the branching ratios are to a good approximation independent of this parameter. Interestingly, the IO regime can feature a situation with a very small branching ratios into tau's, implying approximate zeros in the third column of the Yukawa matrix.}
  \label{tab:BR_Dilep}
\end{table}
From the respective branching ratios one can calculate the suppression factor $B$ of the pair production cross section according to 
\begin{equation}
 B\equiv\sum\limits_{\substack{k=2,3\\ i=e,\mu}} \BR(N_k\ell_i)^2 + \sum\limits_{\substack{k,l=2,3 \\ i,j=e,\mu \\ k\neq l \vee i \neq j}}  \BR(N_k\ell_i)\BR(N_l\ell_j).
\end{equation}
The first sum corresponds to SF production channel, whereas the second term resembles DF as final state particles. We introduced the shorthand notation 
$\BR(N_k\ell_i) \equiv \BR(\sigma \to \ell_i N_k)$. Inserting for instance  the branching ratio for NO (\cref{tab:BR_Dilep}), the final suppression factor is $B\approx 75\%$. While IO yields larger $B$,  we present our results for NO, as we did in \cref{subsec:taus}, since NO is favored by roughly $3\sigma$ from the global fit analyses of neutrino oscillation. Still, we would like to point out that if IO is realized in Nature, not only the sensitivity for this channel would improve, but one would also expect a discovery from neutrinoless double beta experiments which will soon start probing the IO band \cite{Agostini:2018tnm}.\\ 
\begin{figure}[t]
 \centering
 \includegraphics[width=0.5\textwidth]{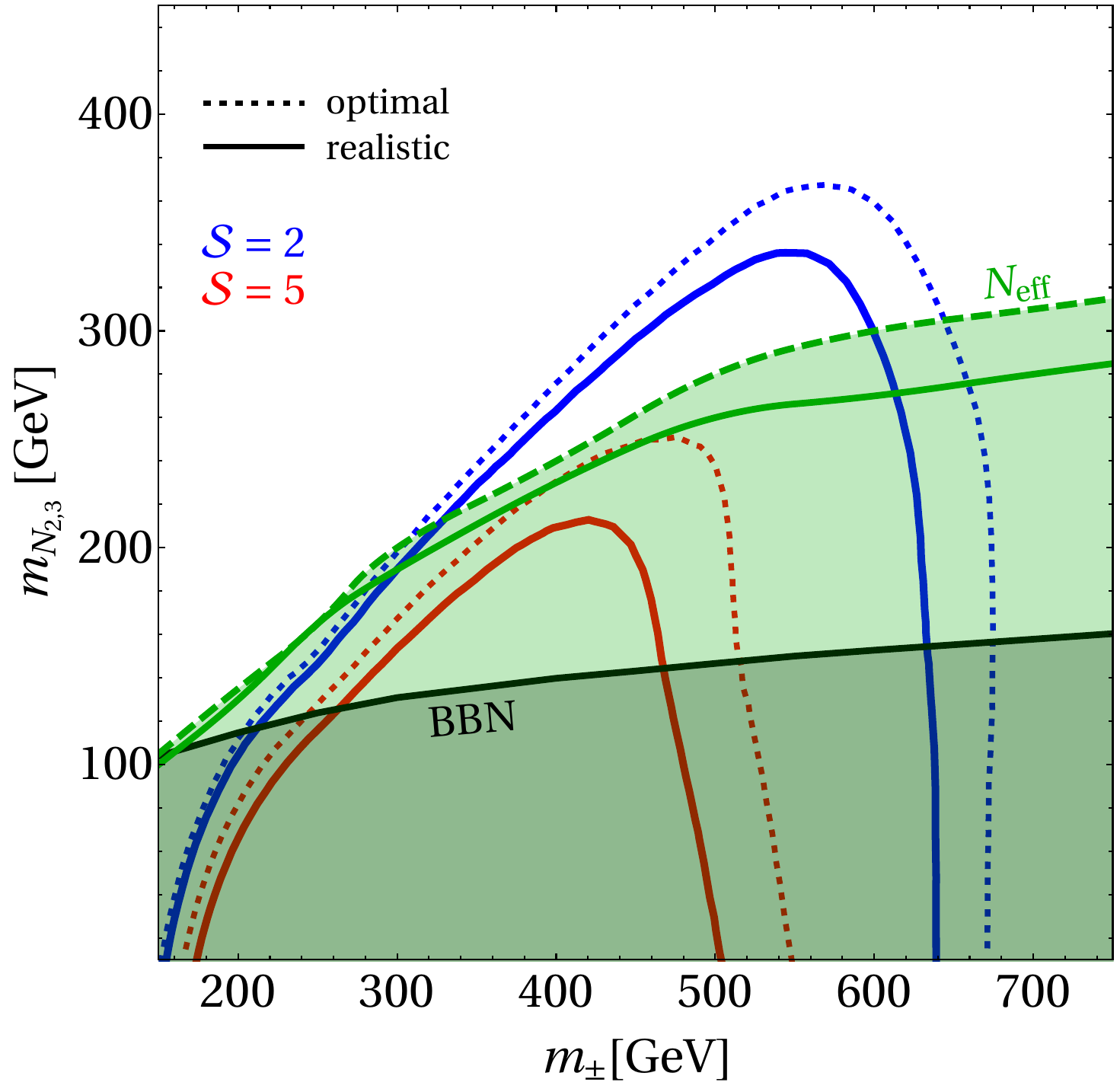}
 \caption{Projected sensitivities for HL-LHC using $\mathcal{L}=4000\invfb$ and the same analysis techniques as \cite{Aaboud:2018jiw}. Given in blue are exclusion limits, $\mathcal{S}=2$, and shown 
 in red are discovery limits, where $\mathcal{S}=5$. The dashed lines correspond to optimal branching ratio into leptons, whereas to obtain the solid lines we used the value for NO given in \cref{tab:BR_Dilep}. HNL masses in the black shaded region are in conflict with BBN constraints discussed in \cref{subsec:BBN} and the green lines arise from $N_\text{eff}$ limits: the dashed green line indicates conservative and the solid one represents aggressive limit. The latter are stronger  for di-lepton searches in contrast with the di-tau case.}
 \label{fig:Dilep_excl}
\end{figure}
The sensitivities are shown in \cref{fig:Dilep_excl} and one can readily infer that, as for the case of di-tau shown in \cref{fig:Ditau_excl}, 
 cosmology disfavors a significant part of the available parameter space. Unfortunately, bounds from $N_\text{eff}$ nearly cover the whole $\mathcal{S}=2$ region, but it is clear that the sensitivity is generally better in comparison with the di-tau\met search.
 

It can be seen by comparing \cref{fig:Ditau_excl,fig:Dilep_excl} that $N_\text{eff}$ limits are stronger for di-lepton case. This is due to the fact, that we can have larger couplings in the di-tau case, because LFV processes yield less stringent bounds on the Yukawa couplings for the third lepton generation. Hence, $N_\text{eff}$ bounds are weakened in such case since larger interaction rates give rise too a smaller freeze-out abundance of $N_2$, suppressing the hot DM component. 

These cosmological bounds starts to flatten for large scalar masses $(m_\pm > 500\GeV)$ and it is therefore tempting to go beyond the energy range of HL-LHC and consider proposed future hadron and lepton colliders. In the following, we will stick to the di-lepton search only, as the results from this section clearly indicate stronger sensitivity in comparison to the di-tau analysis.

\section{Future Colliders}
\label{sec:future}
\noindent
Even though the projections for HL-LHC presented in \cref{sec:pp-collider} do not generally surpass the cosmological bounds, we will demonstrate in this section that future hadron and lepton colliders lead to a different, more promising conclusion.

\subsection{FCC-hh}
\label{subsec:FCC}
\noindent
We start by discussing the scotogenic model in the context of a future circular hadron collider, dubbed FCC-hh \cite{Benedikt:2018csr}. For this purpose, we follow  Ref. \cite{Gori:2014oua}, where an opposite-sign-di-lepton (OSDL) final state with missing energy is discussed with the goal of finding TeV-scale winos and light binos. 

The following variables were used in the analysis to define cuts: $(i)$ $M_\text{eff}$, which is the scalar sum of the $p_T$ of leptons, jets and missing energy (MET)
\begin{equation}
M_\text{eff} = \sum\limits_\text{leptons} p_T + \sum\limits_\text{jets} p_T + \sum\limits_\text{MET} p_T,
\end{equation}
$(ii)$ $M^\prime_\text{eff} = M_\text{eff} - p_T(\ell_1)$, where $p_T(\ell_1)$ is the larger $p_T$ value of the two final state leptons, $(iii)$ the invariant mass of the same sign opposite flavor lepton pair, mSFOS, and, finally, $(iv)$ the transverse mass $M_T$.
We simulated signal events for different mass parameters using the same pipeline as in \cref{sec:pp-collider}. To ensure that our simulations are comparable to those in \cite{Gori:2014oua}, the most dominant backgrounds ($WW$ and $WZ$) were also simulated and compared to the cut flow given in 
\cite{Gori:2014oua}. Our results are presented in \cref{tab:fcc_cut_flow}.\\

\begin{table}[htb]
	\centering
	\begin{tabular}{l@{\hskip 15pt}cc@{\hskip 8pt}cc}
		\toprule 
		Cut & $S$ & $B$ & $\mathcal S_0$ & $\mathcal S_1$ \\
		\midrule
		Baseline & \SI{351}{} & \SI{5.90e5}{} & \SI{0.5}{} & \SI{0.5}{} \\
		$M_{eff}'>\SI{1100}{GeV}$ & \SI{89.7}{} & \SI{625}{} & \SI{3.5}{} & \SI{3.4}{} \\
		$M_T(\mathbf{\cancel{E}_T}\;,l_1+l_2)>\SI{1100}{GeV}$ & \SI{89.7}{} & \SI{234}{} & \SI{5.5}{} & \SI{5.3}{} \\
		$\mathbf{\cancel{E}_T}\;/M_{eff}>0.36$ & \SI{33.2}{} & \SI{62.7}{} & \SI{3.9}{} & \SI{3.6}{} \\
		$p_T(l_2)/p_T(l_1)>0.24$ & \SI{18.6}{} & \SI{18.0}{} & \SI{3.8}{} & \SI{3.4}{} \\
		\bottomrule
	\end{tabular}
\caption{Cuts made for distinguishing signal and background at FCC with a luminosity of \SI{3}{ab^{-1}}. We show the number of signal and background events for $m_{N_{2,3}}=500\GeV$ and $m_\pm = 1\,\text{TeV}$. We employed $\lambda_3=-0.27$ in the analysis. No systematic errors on the background were assumed for this analysis.}
\label{tab:fcc_cut_flow}
\end{table}

Since we generally find that the number of events for signal and background are of the same order, the significance given in \cref{eq:ssb} is not a good approximation and hence for exclusions we use \cref{eq:S1}, while for discoveries we employ \cite{Cowan:2010js}
\begin{align}
\mathcal S_0=\sqrt{2\left((S+B)\log\left(1+\frac SB\right)-S\right)}\,.
\label{eq:S0}
\end{align}

\begin{figure}[t]
	\centering
	\includegraphics[width=.8\textwidth]{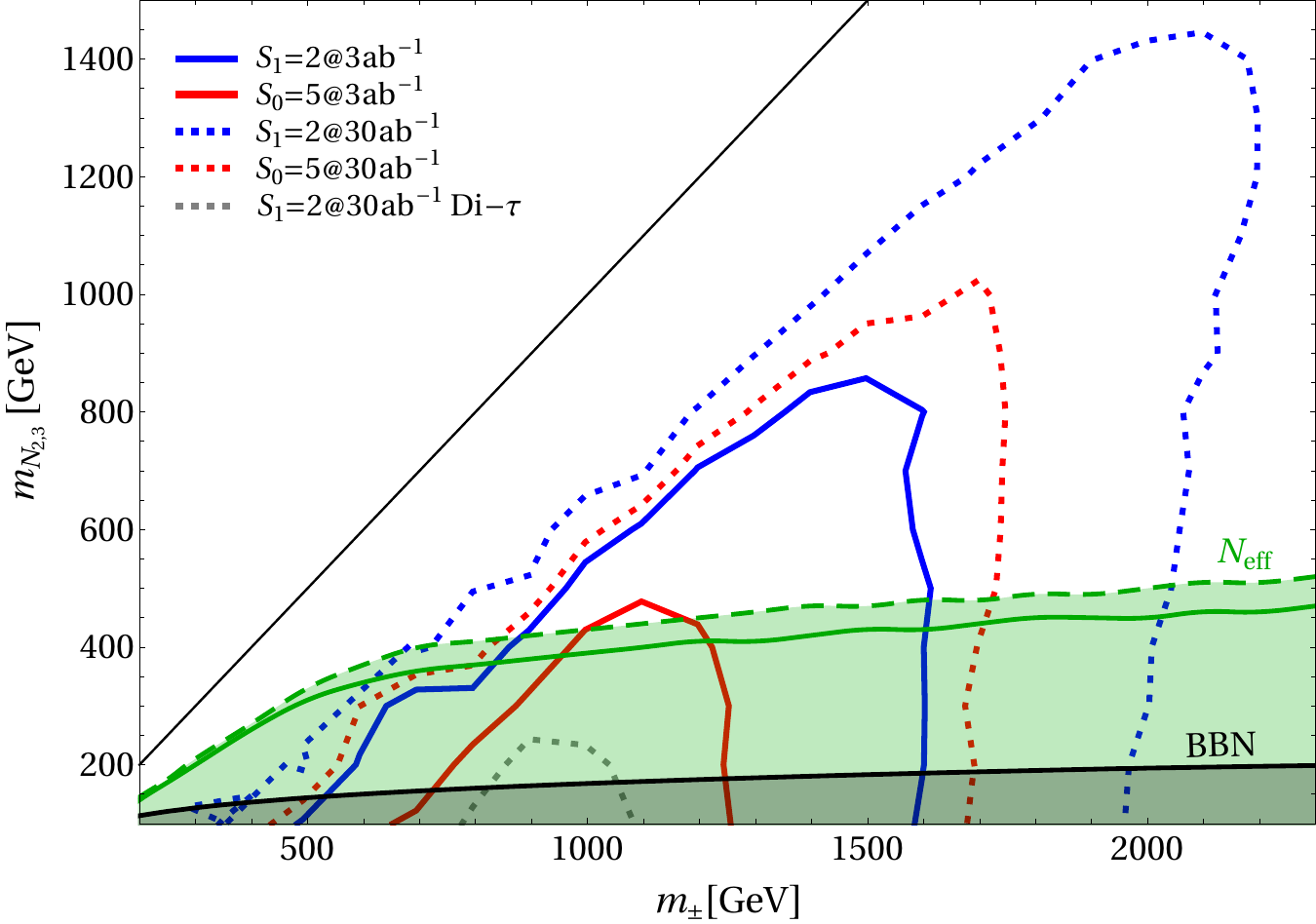}
	\caption{Sensitivity of FCC-hh with $\mathcal L=\SI{3}{ab^{-1}}\;(\SI{30}{ab^{-1}})$ is shown with solid (dashed) lines. The analysis is based on a proposed search for supersymmetry presented in Ref. \cite{Gori:2014oua}. The red and blue curves correspond to the ``best case" scenario with maximized couplings to $e,\,\mu$. While the $\mathcal{S}_0=5$ region for $\mathcal L=\SI{3}{ab^{-1}}$ is in tension with bounds from $N_\text{eff}$, $\mathcal{S}_1=2$ extends to much higher HNL masses, indicating that a significant portion of the parameter space can be probed at FCC-hh. The situation further improves for larger luminosity.
The gray curve corresponds to the ``worst" case in which couplings to tau-lepton are maximized; this case does not feature particularly strong sensitivity. The thin black line corresponds to $m_{N_{2,3}}=m_\pm$.}
	\label{fig:FCC_excl}
\end{figure}

Our findings on the parameter space exclusion capability at FCC-hh with $\mathcal L=\SI{3}{ab^{-1}}$ and $\mathcal L=\SI{30}{ab^{-1}}$ are summarized in \cref{fig:FCC_excl}. Both the contours corresponding to $\mathcal{S}_1=2$ and $\mathcal{S}_0 =5$ are derived for the maximal branching ratios into electrons and muons (see \cref{tab:BR_Dilep}) for normal neutrino mass ordering. This scenario is dubbed ``best case" in \cref{fig:FCC_excl} and it may be inferred that for such couplings FCC-hh will provide the possibility of scanning a large portion of parameter space that is not restricted by BBN and $N_\text{eff}$ constraints. The sensitivity for the ``worst case" scenario in which couplings to $\tau$ lepton are maximized (\cref{tab:BR_Ditau}) is shown in gray in \cref{fig:FCC_excl} and as expected, it is much weaker than the case with dominant couplings to light leptons.

\subsection{CLIC}
\label{subsec:ee-collider}
\noindent
The Compact Linear Collider (CLIC) \cite{deBlas:2018mhx} is a proposed $e^+e^-$ collider that will operate in three stages with  $\sqrt{s}$= \SI{380}{GeV}, \SI{1.5}{TeV}, and \SI{3}{TeV}, respectively \cite{CLIC:2016zwp,deBlas:2018mhx}. In the following, however, we restrict ourselves to the latter case as it offers the  possibility to test larger parameter space of $\mathbf{Z}_2$-odd particle masses in comparison to two other stages. As for FCC-hh, we consider the di-lepton signal.

At $e^+e^-$ colliders there are two complementary processes to produce $\sigma^\pm$ in our model: One is through the exchange of a $Z$ boson or a photon in the $s$-channel; another possibility is via HNLs in the $t$-channel. In the latter case, elements of Yukawa matrix enter in the production cross section. However, we have checked by performing analytic calculations of the cross section that such production is subdominant for the Yukawa couplings employed in this work.

The most outstanding advantage that lepton colliders offer with respect to hadron colliders is the clean signal; without parton distribution functions to be considered, missing energy and momentum can be precisely reconstructed and the distributions in the kinematic variables are not heavily smeared out due to the vastly different energies that interacting partons can have during the collisions. Furthermore, QCD backgrounds are reduced to a degree that allows a more detailed analysis of pure electroweak processes and this is particularly relevant for the model under our consideration.

For general background rejection, we make the following preselection cuts: To get rid of QCD processes, we require the final state to contain no jets\footnote{In particular, we require the number of objects reconstructed as VLCjetR05N2 event types, defined in the official CLIC Delphes-card, to be zero.}. Furthermore, we require exactly two leptons of opposite charge and that \MET exceeds \SI{100}{GeV}. The latter cut suppresses most of the $e^+e^-\to l^+l^-$ background.

The dominant background after the preselection cuts is $e^+e^-\to l^+l^-\nu\bar\nu$ ($l$ stands for any generation of charged leptons) and in the following this is the only background process under our consideration. In the case of $\tau$ leptons in the final state, only those events in which  $\tau$ decays leptonically are considered; in this case there are hence four final state neutrinos. We have checked that any other SM background gives a negligible contribution after the above preselection cuts. A very promising result for the exclusion limits was found by choosing cuts as given in \cref{tab:CLIC_cuts}. 

\begin{table}[t]
 \centering
 \begin{tabular}{c@{\hskip 15pt}cc@{\hskip 8pt}cc@{\hskip 8pt}cc@{\hskip 15pt}}
  \toprule
  & \multicolumn{4}{@{\hskip -15pt}c}{``best case"} \\
  Cut & $S$ & $B$ & $\mathcal S_0$ & $\mathcal S_1$ & $\mathcal S_0'$ & $\mathcal S_1'$ \\
  \midrule
  preselection & $4101$ & $1.0\times10^6$  & $4.1$  & $4.1$ & 1.3 & 1.3   \\
  $M_{eff}>0.5\cdot \mu_2 - 1.2\cdot m_{N_{2,3}} + 1000$ GeV & 3442 &  $2.7\times10^5$  & 6.6  & 6.6 & 3.6 & 3.6 \\
  $\text{mSFOS}>-0.1\cdot \mu_2 - 0.3\cdot m_{N_{2,3}} + 530$ GeV & 3260 & $2.2\times 10^5$  & 6.9  & 6.8 & 4.0 & 4.0  \\
  $|\eta(l_1)|<0.6$ & 2502 & $2.1\times10^4$  & 16.9  & 16.6 & 15.5 & 15.4   \\
  kinematics & 2136 & 908  & 55.6  & 45.6 & 55.3 & 45.5  \\
  \toprule
  & \multicolumn{4}{@{\hskip -15pt}c}{``worst case"} \\
  Cut & $S$ & $B$ & $\mathcal S_0$ & $\mathcal S_1$ & $\mathcal S_0'$ & $\mathcal S_1'$  \\
  \midrule
  preselection & 1188 & $1.0\times10^6$  & 1.2  & 1.2 & 0.4 & 0.4   \\
  $\text{mSFOS}>- 0.3\cdot m_{N_{2,3}} + 130$ GeV & 1153 & $9.7\times10^5$  & 1.2  & 1.2 & 0.4 & 0.4  \\
  $|\eta(l_1)|<0.6$ & 800 & $1.2\times10^5$  & 2.3  & 2.3 & 1.6 & 1.6   \\
  kinematics & 386 & 2806  & 7.1  & 7.0 & 7.0 & 6.9 \\
  \bottomrule
 \end{tabular}
\caption{Cuts made for distinguishing signal and background at CLIC with a center of mass energy of \SI{3}{TeV} and a luminosity of \SI{5}{ab^{-1}}. We show the number of signal and background events together with the corresponding sensitivities for a benchmark point $m_{N_{2,3}}=500\GeV$ and $m_\pm = 1\,\text{TeV}$. The results are shown both for ``best" and ``worst" case scenarios which correspond to maximizing couplings to e,$\mu$ and $\tau$ lepton, respectively. Details about the systematic errors are given in the text.}
\label{tab:CLIC_cuts} 
\end{table}

Regarding cuts, the variables mSFOS and $M_\text{eff}$, already employed for our FCC-hh analysis, were reused for CLIC because of their great potential for discriminating the scotogenic model, with its comparatively large masses of $\mathbf{Z}_2$-odd particles, against the SM. Furthermore, the pseudorapidity of the first lepton, $\eta(l_1)$, turns out to be a very useful variable; it peaks at large values for the SM background while most of the signal events are more central in this variable. 

Regarding the large number of events still present after the final cuts, systematic uncertainties have to be considered, as they more dominantly affect the significance in this case. For that reason, the significances in \cref{eq:S0} for discovery and in \cref{eq:S1} for exclusion have to be adapted to include these systematic uncertainties. Assuming the systematic uncertainties to be described by Gaussian noise with standard deviation $\sigma_B=\varepsilon B$, one finds
\begin{align}
 \mathcal S_0'=&\sqrt{2(s+b)\ln\left(\frac{2(s+b)}{b+a_0-\sigma_b^2}\right)-s+\frac{\left(b-\sigma_b^2\right)\left(b-a_0+\sigma_b^2\right)}{2\sigma_b^2}}, \\
 \mathcal S_1'=&\sqrt{a_1+s-b-\sigma_b^2+\frac{\left(s+b-a_1+\sigma_b^2\right)^2}{4\sigma_b^2}+2b\ln\left(\frac{2b}{s+b+a_1-\sigma_b^2}\right)},
\end{align}
where
\begin{align*}
 a_0=&\sqrt{b^2+2(2s+b)\sigma_b^2+\sigma_b^4}, \\
 a_1=&\sqrt{(s+b)^2+2(b-s)\sigma_b^2+\sigma_b^4}.
\end{align*}
We followed \cite{deBlas:2018mhx} and assumed a background uncertainty of $0.3\%$ in the following analysis.

A further suppression of the background can be made by utilizing the kinematics of this process. In the di-lepton search at CLIC, one can reconstruct the  4-momenta of the two HNLs (denoted by $p_{3,4}^\mu$ in what follows) by using only the initial state energy $\sqrt{s}$  and the known 4-momenta of the two charged leptons ($p_{1,2}^\mu$). Four out of these eight unknowns are fixed by 4-momentum conservation, namely $\sum_{i=1}^4 p_i^\mu=\left(\sqrt{s}\,,0\,,0\,,0\right)$. Furthermore, imposing that all  intermediate and final state particles are on-shell yields (up to combinatorics) the following four relations:
\begin{align}
&(p_1+p_3)^2 = m_\pm^2=(p_2+p_4)^2\,, &  p_3^2 = p_4^2 = m_{N_{2,3}}^2\,.
\label{eq:on-shell}
\end{align}

 In total, we end up with a solvable system of equations. For separating signal and background events we use that the latter ones typically have
different kinematic properties, since the mass of the intermediate particle is for instance set by the $W$ boson mass. After requiring that $(i)$ there is a physical solution to the aforementioned equations, i.e. we demand the momenta of the invisible particles to be real valued, and $(ii)$ there is no real valued solution if $m_\pm^2$ and $m_{N_{2,3}}^2$ in \cref{eq:on-shell} get replaced with the W boson and active neutrino mass, respectively, we reach the following effect: The number of signal events typically decreases only by 70\%, whereas background is strongly reduced, at most only a few percent of such events survive this cut. In reality, the 4-momenta in \cref{eq:on-shell} are only known to a finite precision. So it is crucial to include the finite detector resolution by using the respective \texttt{Delphes} card for the CLIC detector which is based on \cite{Arominski:2018uuz}. As such, the simulated $p_T$ values experience a smearing with respect to the detector resolution.
 
\begin{figure}[t]
	\centering
	\includegraphics[width=.8\textwidth]{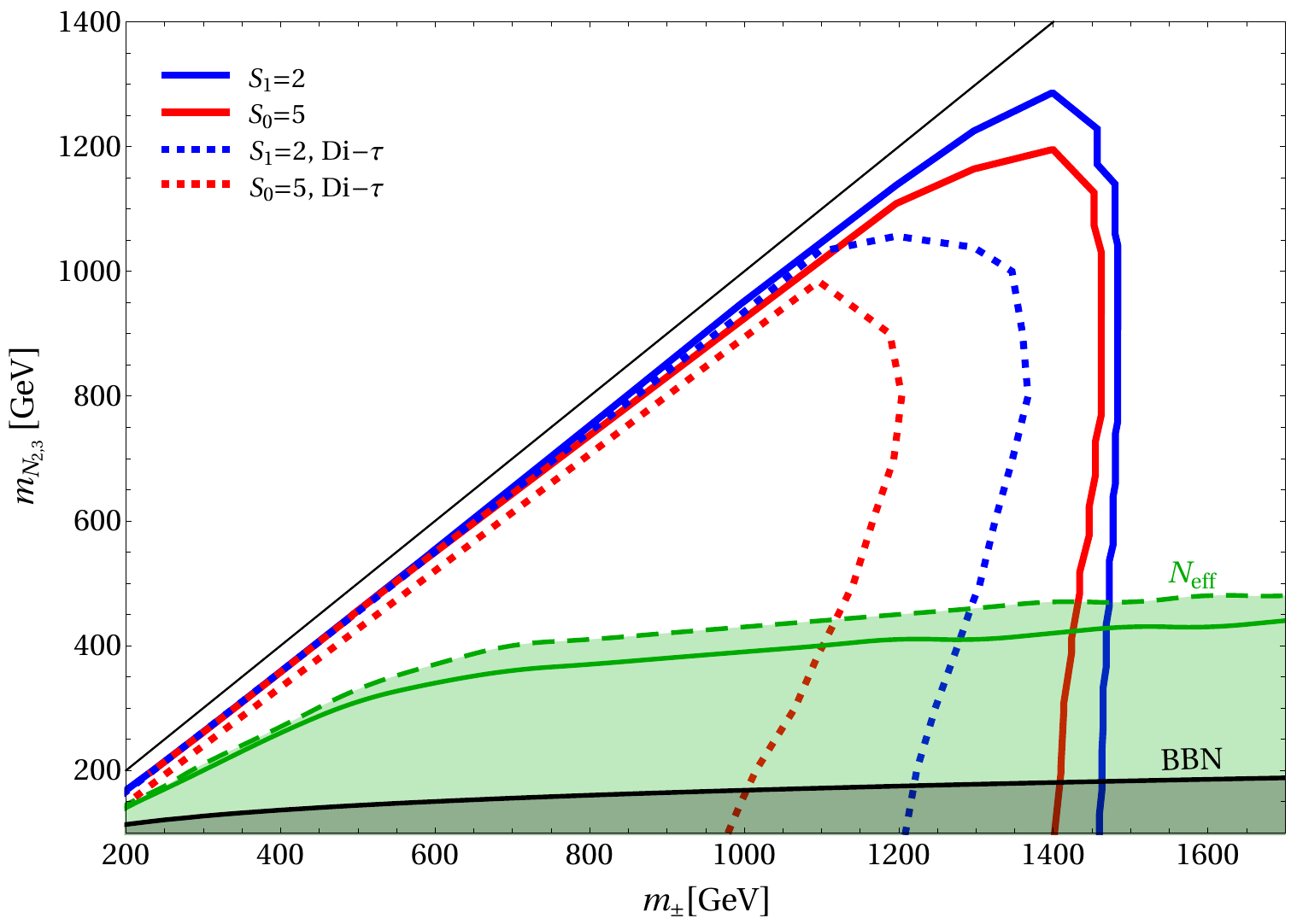}
	\caption{CLIC sensitivity for the di-lepton search. Using maximized couplings to $e,\,\mu$ we obtained the red solid contour that corresponds to a $5\sigma$ discovery and the blue one that represents $2\sigma$ exclusion. The corresponding dashed contours are for  the case where $\tau$ couplings are maximized. The thin black line indicates $m_{N_{2,3}}=m_\pm$. We assumed an background uncertainty of $0.3\%$.}
	\label{fig:CLIC_exclusion}
\end{figure}

We wish to stress that the cuts on  mSFOS and $M_\text{eff}$ are not optimal for each parameter point since we only used a very simple function of the model parameters. Still, the resulting exclusion limits, shown in \cref{fig:CLIC_exclusion}, already indicate the great potential for testing this model setup at CLIC; sensitivity curves both for the ``best case" and the ``worst case" scenario  exceed $N_\text{eff}$ limits. Interestingly, $\mathcal{S}_1=2$ and $\mathcal{S}_0=5$ curves reach similar limits for these two cases. This is a consequence of the aforementioned use of kinematics: Due to the suppression of background events, we find large sensitivity values across the parameter space. Although the ``worst case" features fewer event rates due to smaller branching ratios, there is still a sizable sensitivity available.\\

In comparison with FCC sensitivities at $3\,\text{ab}^{-1}$, CLIC can reach higher HNL masses while being less sensitive to $\sigma^\pm$ masses larger than \SI{1400}{GeV}; this is to be expected with a total center of mass energy of \SI{3000}{GeV}. Overall, CLIC offers a testable parameter space which is comparable to the FCC result. While the reach at CLIC is not as large in comparison to FCC with $30\,\text{ab}^{-1}$, we note that, unlike FCC, CLIC can nearly close the available kinematic window, having the sensitivity in the vicinity of $m_{N_{2,3}}=m_\pm$ line. Let us note that this is the kinematic window in which coannihilations are very effective and this can strongly suppress the hot DM component, relaxing the $N_\text{eff}$ limits, while BBN limits still play a role. We wish to stress that such small splitting between $\mathbf{Z}_2$-odd fermions and scalars and $m_{N_{2,3}}\simeq 1$ TeV is exactly the setup in which we have previously \cite{Baumholzer:2018sfb} shown that the observed amounts of DM and baryon asymmetry of the Universe can be simultaneously explained within the scotogenic model. It is very intriguing that future lepton colliders will have a sensitivity to such a scenario.

In conclusion, we have shown in this section that both lepton and hadron colliders offer promising and complementary ways to look for the considered scotogenic scenario.

\subsection{Summary of collider searches}
\label{subsec:summary_collider}
\begin{figure}[t]
	\centering
	\includegraphics[width=.8\textwidth]{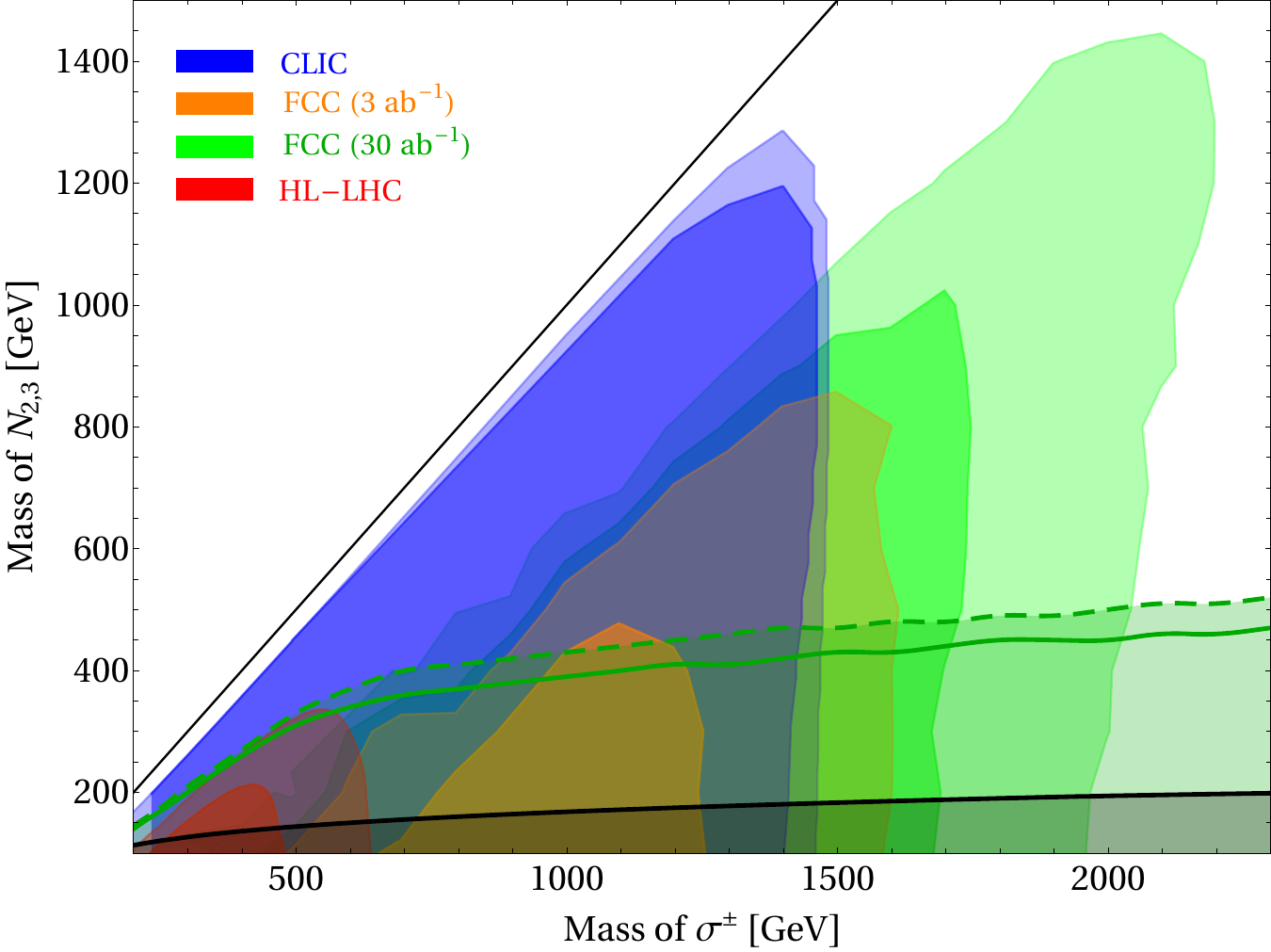}
	\caption{Summarized sensitivity curves as discussed in \cref{subsec:leptons,subsec:ee-collider,subsec:FCC}. The light shaded regions correspond to $S=2$
	and the darker regions represent $S=5$. Again, the thin black line indicates $m_{N_{2,3}}=m_\pm$ whereas thick black line shows BBN constraints; green solid (dashed) curves represent \textit{conservative} (\textit{aggressive}) $N_\text{eff}$ constraints. We assumed a maximal coupling to $e,\,\mu$ in this case.}
	\label{fig:summary}
\end{figure}

\noindent
Having explored the capability of HL-LHC, as well as future hadron (FCC-hh) and lepton (CLIC) colliders for testing the scotogenic model, we summarize the situation in \cref{fig:summary}. The figure contains sensitivity curves as well as BBN and $N_\text{eff}$ limits already presented in \cref{fig:Ditau_excl,fig:Dilep_excl,fig:FCC_excl,fig:CLIC_exclusion}. While the discovery at HL-LHC is less likely due to the tension with cosmological limits, the future colliders offer more promising situation in which large portions of the parameter space can be tested.\\

While we have shown that the future collider prospects are bright, it is fair to note that the cosmological and other terrestrial searches will also provide stronger limits, particularly for heavier HNLs. In what follows we list most notable among these projects: 

\begin{itemize}
	\item New searches for LFV processes can lower the bounds on Yukawa couplings $y_{2\alpha}$ and $y_{3,\alpha}$; for instance MEG II \cite{Baldini:2018nnn} features a projected sensitivity of $\BR (\mu \to e\,\gamma) < 6\cdot 10^{-14}$; an improvement of about an order of magnitude compared to the previous bound.
	\item New observations of small scale structure in combination with detailed simulations of warm DM will push $m_\text{NRP}$ to larger values and this in turn requires smaller $y_{1\alpha}$ to produce the observed DM abundance via freeze-in, which will also impact BBN limits.
	\item Upcoming CMB experiments will measure $N_\text{eff}$ to a precision of $\Delta N_\text{eff} = 0.06$ \cite{Abazajian:2019eic} which leaves less room for a hot DM subcomponent.
\end{itemize}

To summarize, these rather complementary searches would probe the parameter space up to even smaller mass ratios between $\sigma^\pm$ and HNLs. They directly or indirectly set a stricter upper limit on the abundance of $N_2$ which is crucial for cosmology. 
Hence, in the near future these experiments will offer novel relations between collider searches and cosmological observations.

\section{Summary and Conclusions}
\label{sec:conclusion}
\noindent
The scotogenic model is a very popular extension of the Standard Model which can explain both neutrino masses and the origin of DM. 
In this paper we focused on the case of keV-scale fermionic DM  with the mass of remaining $\mathbf{Z}_2$-odd fermion and scalar degrees of freedom at $\mathcal{O}(100)\GeV$. In this setup there are two distinct DM production mechanisms: freeze-in through the decays of heavy scalars and the production from the decays of next-to-lightest $\mathbf{Z}_2$-odd particle, $N_2$, produced via freeze-out. The large mass gap between DM and $N_2$ can generally allow for a sufficient suppression of the abundance arising from the latter mechanism, which is required as the corresponding DM momentum distribution is hot and could hence lead to washout of structures at small scales. We have shown that even stronger constraints arise from the contribution of such hot DM to $N_\text{eff}$. We also derived BBN bounds from the requirement that $N_2$ particles decay within $\sim 1$ second. 

Armed with the limits from cosmology, we focused on collider phenomenology; in particular we studied  $p\,p \to \sigma^\pm \, \sigma^\mp \to \ell^\pm\, \ell^\mp \, +\,\cancel{E_T}$ channels. We have demonstrated that testing this model at colliders strongly relies on forthcoming stages of LHC as well as future colliders since we were not able to extract robust bounds by using $36.1\invfb$ data. Thus, we further considered HL-LHC, FCC-hh and CLIC.
 For the former, we found that the testable region is in tension with $N_\text{eff}$ and BBN limits, whereas both CLIC and FCC can probe significant regions of parameter space free from cosmological bounds.
The sensitivity reach of CLIC exceeds 1 TeV for both HNL and charged scalar masses. The FCC with $3\invab$ would reach similar scalar and even larger HNL masses, and this further improves for $30\invab$ where HNL masses up to 2 TeV would be probed.

 
In summary, we have shown that if Nature has chosen the scotogenic scenario, there is a number of complementary tests, stemming from terrestrial experiments to cosmological surveys 
which indicate that the model has a rich phenomenology and a very promising discovery potential.


\tocless{\section*{Acknowledgements}}
\noindent
SB would like to thank the theory division at Fermilab for the hospitality during the final stage of the project and Miguel Escudero and Sam Witte for fruitful discussions regarding the calculation of $N_\text{eff}$. VB would like to thank Thomas Hugle for useful discussions. 
Research in Mainz is supported by the Cluster of Excellence “Precision Physics, Fundamental Interactions, and Structure of Matter” (PRISMA+ EXC 2118/1) funded by the German Research Foundation(DFG) within the German Excellence Strategy (Project ID 39083149), and by grant 05H18UMCA1 of the German Federal Ministry for Education and Research (BMBF). 

\bibliographystyle{JHEP}
\tocless\bibliography{scotogenic}

\end{document}